\documentclass[prd,twocolumn,floatfix,amsmath,nofootinbib,amssymb,floatfix]{revtex4}
\usepackage{graphicx,color,dcolumn,booktabs,bm}
\usepackage{longtable,lscape}
\usepackage{pdfpages}
\usepackage{txfonts}
\usepackage{overpic}
\usepackage{amssymb}
\usepackage{makecell}
\usepackage{indentfirst}
\usepackage{feynmf}   %{feynmp}
\usepackage{slashed}  %for Feynman symbols
\usepackage{cases}
\usepackage{color}
\usepackage{multirow}
\usepackage{threeparttable}
\usepackage{epstopdf}
\usepackage{enumerate}
\usepackage{subfigure}
\usepackage{diagbox}
\usepackage{graphicx,color,dcolumn,booktabs,bm}
\usepackage{mathrsfs}
\usepackage{cancel}
\usepackage{float}
\usepackage[misc]{ifsym}
\usepackage[colorlinks,
            citecolor=blue,
            anchorcolor=red,
            menucolor=red,
            linkcolor=red,
            filecolor=red,
            runcolor=red,
            urlcolor=blue,
            frenchlinks=red]{hyperref}
\usepackage{array}
\usepackage{booktabs}

\begin{document}
\title{Analysis of the semileptonic decays $\Lambda_b\to\Lambda_cl\bar{\nu}_l$ and $\Xi_b\to\Xi_cl\bar{\nu}_l$ in QCD sum rules}
\author{Jie Lu$^{1}$}
\email{l17693567997@163.com}
\author{Guo-Liang Yu$^{2}$}
\email{yuguoliang2011@163.com}
\author{Dian-Yong Chen$^{1,3}$}
\email{chendy@seu.edu.cn}
\author{Zhi-Gang Wang$^{2}$}
\email{zgwang@aliyun.com}
\author{Bin Wu$^{1}$}

\affiliation{$^1$ School of Physics, Southeast University, Nanjing 210094, People's Republic of
China\\$^2$ Department of Mathematics and Physics, North China
Electric Power University, Baoding 071003, People's Republic of
China\\$^3$ Lanzhou Center for Theoretical Physics, Lanzhou University, Lanzhou 730000, People's Republic of
China}
\date{\today }

\begin{abstract}
In this article, the electroweak transition form factors of $\Lambda_b\to\Lambda_c$ and $\Xi_b\to\Xi_c$ are analyzed within the framework of three-point QCD sum rules. In phenomenological side, all possible couplings of interpolating current to hadronic states are considered. In QCD side, the perturbative part and the contributions of vacuum condensates up to dimension 8 are also included. With the estimated form factors, we study the decay widths and branching ratios of semileptonic decays $\Lambda_b\to\Lambda_cl\bar{\nu}_l$ and $\Xi_b\to\Xi_cl\bar{\nu}_l$ ($l=e,\mu$ and $\tau$). Our results for the branching ratios of $\Lambda_b\to\Lambda_cl\bar{\nu}_l$ are comparable with experimental data and the results from other collaborations. In addition, our prediction for the branching ratio of $\Xi_b\to\Xi_cl\bar{\nu}_l$ can provide a valuable reference for future experimental measurements.
\end{abstract}


\maketitle

\section{Introduction}\label{sec1}
The study of heavy flavored baryons is of great significance in both theory and experiment as it can provide a suitable place to test the standard model (SM) and find the new physics beyond the SM. Among these, the semileptonic decays $\Lambda_b\to\Lambda_cl\bar{\nu}_l$ and $\Xi_b\to\Xi_cl\bar{\nu}_l$ can be used to extract the Cabibbo-Kobayashi-Maskawa (CKM) matrix element $V_{cb}$. Furthermore, those processes also play an important role to test the lepton universality.

In terms of experiments, the measured branching ratio for the semileptonic decays of $\Lambda_b$ are given as~\cite{ParticleDataGroup:2024cfk}:
\begin{eqnarray}
\notag
&&\mathcal{B}[\Lambda_b\to\Lambda_c(e,\mu)\bar{\nu}_{e,\mu}]=6.2^{+1.4}_{-1.3}\%,\\
&&\mathcal{B}[\Lambda_b\to\Lambda_c\tau\bar{\nu}_\tau]=1.9\pm0.5\%,
\end{eqnarray}
and the lepton universality ratio $R_{\Lambda_c}=\mathcal{B}[\Lambda_b\to\Lambda_c\tau\bar{\nu}_\tau]/\mathcal{B}[\Lambda_b\to\Lambda_c\mu\bar{\nu}_\mu]$ is also reported by LHCb Collaboration as~\cite{LHCb:2017vhq,LHCb:2022piu}:
\begin{eqnarray}
R_{\Lambda_c}=0.242\pm0.026\pm0.040\pm0.059,
\end{eqnarray}
where the first uncertainties are systematic, the second and third are originated from statistical in $\mathcal{B}[\Lambda_b\to\Lambda_c\tau\bar{\nu}_\tau]$ and $\mathcal{B}[\Lambda_b\to\Lambda_c\mu\bar{\nu}_\mu]$, respectively. For $\Xi_b$ baryon, the semileptonic decay branching ratio and $R_{\Xi_c}$ have not been determined yet experimentally. However, the LHCb Collaboration recently observed the decays $\Xi_b^0\to \Xi_c^+D_s^-$ and $\Xi_b^-\to \Xi_c^0D_s^-$~\cite{LHCb:2023ngz}, which indicates experimental measurement of $\Xi_b\to \Xi_c$ transition may be promising in the near future.

On the theoretical aspect, the semileptonic decays $\Lambda_b\to\Lambda_cl\bar{\nu}_l$ and $\Xi_b\to \Xi_cl\bar{\nu}_l$ are both dominated by the decay process $b\to cl\bar{\nu}_l$ at the quark level, and can be described by the electroweak effective Hamiltonian in SM. However, these decay processes involve both electroweak and strong interactions, and are difficult to study with perturbation methods. Therefore, some non-perturbation approaches are employed to carry out this work such as the Lattice QCD (LQCD)~\cite{Detmold:2015aaa}, the QCD sum rules (QCDSR)~\cite{Azizi:2018axf,Zhao:2020mod,Neishabouri:2025abl}, Light-cone QCD sum rules (LCSR)~\cite{Miao:2022bga,Duan:2022uzm}, various quark models~\cite{Ebert:2006rp,Gutsche:2014zna,Gutsche:2015mxa,Faustov:2016pal,Faustov:2018ahb,Dutta:2018zqp,Zhao:2018zcb,Zhu:2018jet,Thakkar:2020vpv,Li:2021qod} and others~\cite{Ivanov:1998ya,Shih:1999yh,Dutta:2015ueb,Zhang:2022bvl,Rui:2025iwa,Li:2025rsm}.

As a QCD-based approach to deal with hadronic parameters, the QCDSR is time-honored and widely used in studying the properties of hadrons~\cite{Shifman:1978bx,Shifman:1978by,Colangelo:2000dp,Wang:2025sic}. Specifically, the QCDSR based on three-point correlation function is used to analyze the hadron transition form factors~\cite{Wang:2007ys,Wang:2008pq,Shi:2019hbf,Xing:2021enr,Zhao:2021sje,Zhang:2023nxl,Neishabouri:2024gbc,Tousi:2024usi,Lu:2024tgy,Wu:2024gcq,Lu:2025bvi} and the coupling constants of strong vertices~\cite{Bracco:2011pg,Azizi:2014bua,Rodrigues:2017qsm,Yu:2019sqp,Lu:2023lvu,Lu:2023pcg}. Although, similar works for the semileptonic decays of $\Lambda_b\to\Lambda_cl\bar{\nu}_l$ and $\Xi_b\to\Xi_cl\bar{\nu}_l$ have been already carried out by other collaborations in recent years. More efforts should be made to improve the reliability of the final results. Specifically, the coupling of positive parity baryon interpolating current to negative parity baryon is neglected in Refs.~\cite{Azizi:2018axf,Neishabouri:2025abl}. This leads to dependence of the form factors on the selection of Dirac structures. However, the authors did not show in their analysis the effects of different Dirac structures on the form factors. In addition, the authors considered all possible couplings of interpolating current to hadronic states in phenomenological side in Ref.~\cite{Zhao:2020mod}, but the contributions from higher dimension vacuum condensate such as four quark condensate are neglected which play important roles in non-perturbative contributions. In the present work, we analyze the electroweak transition form factors of $\Lambda_b\to\Lambda_c$ and $\Xi_b\to\Xi_c$ using the three-point QCDSR, where all possible couplings of interpolating current to hadronic states are considered and the operator product expansion (OPE) is truncated at dimension of 8. Then, we also analyze the decay widths and branching ratios of semileptonic decays $\Lambda_b\to\Lambda_cl\bar{\nu}_l$ and $\Xi_b\to\Xi_cl\bar{\nu}_l$ with the calculated form factors.

This work is organized as follows. After introduction in Sec. \ref{sec1}, the semileptonic decay processes $\Lambda_b\to\Lambda_cl\bar{\nu}_l$ and $\Xi_b\to\Xi_cl\bar{\nu}_l$ are analyzed in Sec. \ref{sec2}, and the electroweak transition form factors are introduced. In Sec. \ref{sec3}, the electroweak transition form factors of $\Lambda_b\to\Lambda_c$ and $\Xi_b\to\Xi_c$ are analyzed within the three-point QCDSR. Sec. \ref{sec4} is employed to present the numerical results and discussions and Sec. \ref{sec5} is the conclusion part.
\section{Semileptonic decays of $\Lambda_b\to\Lambda_cl\bar{\nu}_l$ and $\Xi_b\to\Xi_cl\bar{\nu}_l$}\label{sec2}
\begin{figure}
	\centering
	\includegraphics[width=8.5cm]{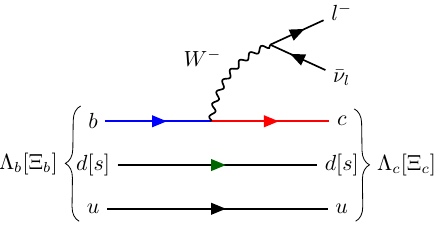}
	\caption{The Feynman diagram of semi-leptonic decay processes $\Lambda_b[\Xi_b]\to \Lambda_c[\Xi_b]l\bar{\nu}_{l}$.}
	\label{FDH}
\end{figure}

The semileptonic decays $\Lambda_b\to\Lambda_cl\bar{\nu}_l$ and $\Xi_b\to\Xi_cl\bar{\nu}_l$ are both dominated by the transition $b\to cl\bar{\nu}_l$ at the quark level. The corresponding effective Hamiltonian can be written as: 
\begin{eqnarray}\label{eq:1}
H_{eff} = \frac{G_F}{\sqrt 2 }V_{cb}\bar c\gamma _\mu (1 - \gamma _5)b\bar{ v}_l\gamma _\mu (1 - \gamma _5)l,
\end{eqnarray}
where $G_F=1.16637\times10^{-5}$ GeV$^{-2}$ is the Fermi constant and $V_{cb}$ is the CKM matrix element. The corresponding Feynman diagram is shown as Fig. \ref{FDH}.

With above Hamiltonian, the transition matrix element of these decay processes can be written as:
\begin{eqnarray}\label{eq:2}
\notag
T &&= \left\langle \Lambda _c[\Xi_c]l\bar{\nu}_l\right|H_{eff}\left| \Lambda _b[\Xi_b] \right\rangle \\
\notag
&&=\frac{G_F}{\sqrt 2}V_{cb}\left\langle \Lambda _c[\Xi_c]\right|\bar c\gamma _\mu (1 - \gamma _5)b\left| \Lambda _b[\Xi_b] \right\rangle\\
&&\times \left\langle l\bar{\nu}_l\right|\bar v_l\gamma _\mu(1 - \gamma _5)l\left| 0 \right\rangle,
\end{eqnarray}
where the leptonic matrix element can be formulated as the following form by electroweak perturbation theory
\begin{eqnarray}\label{eq:3}
\langle l\bar{\nu}_l |\bar v_l\gamma _\mu(1 - \gamma _5)l | 0 \rangle  = \bar u_{v,s}\gamma _\mu(1 - \gamma _5)u_{ - l,s'},
\end{eqnarray}
Here the $u_{v,s}$ and $u_{-l,s'}$ are the spinor wave functions of $\bar{\nu}_l$ and $l$, and the subscripts $v(l)$ and $s(s')$ represent the four momentum and spin.

The hadronic electroweak transition matrix element can not be calculated perturbatively due to non-perturbation effect of low energy QCD. In general, this matrix element can be parameterized as following form factors
\begin{widetext}
\begin{eqnarray}\label{eq:4}
\notag
\left\langle \Lambda _c[\Xi _c] \right|\bar c\gamma _\mu(1 - \gamma _5)b \left| \Lambda _b[\Xi _b] \right\rangle &&= \bar u(p',s')\left[ \gamma _\mu f_1(q^2) + i\frac{\sigma _{\mu \nu }q^\nu }{m_{\Lambda _b[\Xi _b]}}f_2(q^2) + \frac{q_\mu }{m_{\Lambda _b[\Xi _b]}}f_3(q^2) \right]U(p,s)\\
&&- \bar u(p',s')\left[ \gamma _\mu g_1(q^2) + i\frac{\sigma _{\mu \nu }q^\nu }{m_{\Lambda _b[\Xi _b]}}g_2(q^2) + \frac{q_\mu}{m_{\Lambda _b[\Xi _b]}}g_3(q^2) \right]\gamma _5U(p,s),
\end{eqnarray}
\end{widetext}
where $q=p-p'$, $\sigma_{\mu\nu}=\frac{i}{2}[\gamma_\mu,\gamma_\nu]$. $U(p,s)$ and $u(p',s')$ denote the spinor wave functions of $\Lambda_b[\Xi_b]$ and $\Lambda_c[\Xi_c]$. $f_i(q^2)$ and $g_i(q^2)$ $(i=1,2,3)$ are the vector and axial vector form factors, respectively.
\section{Three-point QCD sum rules for transition form factors}\label{sec3}
To calculate the electroweak transition form factors by QCDSR, the following three-point correction function is firstly constructed:
\begin{eqnarray}\label{eq:5}
\notag
\Pi _\mu(p,p') &&= i^2\int d^4x d^4ye^{ip'x}e^{iqy}\\
&&\times \left\langle 0 \right|\mathcal{T}[J_{\Lambda _c[\Xi_c]}(x)J_\mu ^{V - A}(y)\bar J_{\Lambda _b[\Xi_b]}(0)]\left| 0 \right\rangle, 
\end{eqnarray}
where $\mathcal{T}$ denotes the time ordered product and $\bar{J}=J^\dagger\gamma_0$. $J_{\Lambda_b[\Xi_b]}$ and $J_{\Lambda_c[\Xi_c]}$ are the interpolating currents of $\Lambda_b[\Xi_b]$ and $\Lambda_c[\Xi_c]$, respectively. $J^{V-A}_\mu$ is the electroweak transition current. These currents are taken as the following forms:
\begin{eqnarray}\label{eq:6}
\notag
J_{\Lambda _c[\Xi _c]}(x) &&= \varepsilon _{ijk}\left( u^{iT}(x)\mathcal{C}\gamma _5d[s]^j(x) \right)c^k(x),\\
\notag
J_\mu ^{V - A}(y) &&= \bar c(y)\gamma _\mu(1 - \gamma _5)b(y),\\
J_{\Lambda _b[\Xi _b]}(0) &&= \varepsilon _{i'j'k'}\left( u^{i'T}(0)\mathcal{C}\gamma _5d[s]^{j'}(0) \right)b^{k'}(0),
\end{eqnarray}
where $\varepsilon_{ijk}$ is the 3 dimension Levi-Civita tensor, $i$, $j$ and $k$
represent the color indices, and $\mathcal{C}$ is the charge conjugation operator.

In the framework of QCDSR, the above three-point correlation function can be calculated at both hadron and quark levels, which are called the phenomenological side and the QCD side, respectively. Finally, the sum rules for the form factors are obtained by matching the calculation of these two levels and using the quark-hadron duality condition.
\subsection{The phenomenological side}\label{sec3.1}
In the phenomenological side, the complete set of
hadronic states coupled to corresponding interpolating currents is inserted into the correlation function. After performing the integrals in coordinate space, isolating the contributions of the ground states from excited states and using the double dispersion relation, the three-point correlation function can be represented as the following form~\cite{Colangelo:2000dp}:
\begin{widetext}
\begin{eqnarray}\label{eq:7}
\notag
\Pi _\mu ^{\mathrm{phy}}(p,p') &&= \frac{\left\langle 0 \right|J_{\Lambda _c[\Xi _c]}(0) \left| \Lambda _c[\Xi _c](p') \right\rangle \left\langle \Lambda _c[\Xi _c](p') \right|J_\mu ^{V - A}(0)\left| \Lambda _b[\Xi _b](p)\right\rangle \left\langle \Lambda _b[\Xi _b](p) \right|\bar J_{\Lambda _b[\Xi _b]}(0) \left| 0 \right\rangle}{(m_{\Lambda _c[\Xi _c]}^2-p'^2)(m_{\Lambda _b[\Xi _b]}^2-p^2)}\\
\notag
&&+ \frac{\left\langle 0 \right|J_{\Lambda _c[\Xi _c]}(0)\left| \Lambda _c^*[\Xi _c^*](p') \right\rangle \left\langle \Lambda _c^*[\Xi _c^*](p') \right|J_\mu ^{V - A}(0)\left| \Lambda _b[\Xi _b](p) \right\rangle \left\langle \Lambda _b[\Xi _b](p) \right|\bar J_{\Lambda _b[\Xi _b]}(0)\left| 0 \right\rangle }{(m_{\Lambda _c^*[\Xi _c^*]}^2-p'^2)(m_{\Lambda _b[\Xi _b]}^2-p^2)}\\
\notag
&&+ \frac{\left\langle 0 \right|J_{\Lambda _c[\Xi _c]}(0)\left| \Lambda _c[\Xi _c](p') \right\rangle \left\langle \Lambda _c[\Xi _c](p') \right|J_\mu ^{V - A}(0)\left| \Lambda _b^*[\Xi _b^*](p) \right\rangle \left\langle \Lambda _b^*[\Xi _b^*](p) \right|\bar J_{\Lambda _b[\Xi _b]}(0)\left| 0 \right\rangle }{(m_{\Lambda _c[\Xi _c]}^2-p'^2)(m_{\Lambda _b^*[\Xi _b^*]}^2-p^2)}\\
&&+ \frac{\left\langle 0 \right|J_{\Lambda _c[\Xi _c]}(0)\left| \Lambda _c^*[\Xi _c^*](p') \right\rangle \left\langle \Lambda _c^*[\Xi _c^*](p') \right|J_\mu ^{V - A}(0)\left|\Lambda _b^*[\Xi _b^*](p)\right\rangle \left\langle \Lambda _b^*[\Xi _b^*](p) \right|\bar J_{\Lambda _b[\Xi _b]}(0)\left| 0 \right\rangle}{(m_{\Lambda _c^*[\Xi _c^*]}^2-p'^2)(m_{\Lambda _b^*[\Xi _b^*]}^2-p^2)} + ...,
\end{eqnarray}
where the ellipsis denotes the contributions for higher resonances and continuum states of hadrons. It can be seen from Eq.~(\ref{eq:7}) the interpolating currents $J_{\Lambda_b[\Xi_b]}$ and $J_{\Lambda_c[\Xi_c]}$ can not only couple to $\Lambda_b[\Xi_b]$ and $\Lambda_c[\Xi_c]$ with spin parity $\frac{1}{2}^+$, but also couple to $\Lambda^*_b[\Xi^*_b]$ and $\Lambda^*_c[\Xi^*_c]$ with spin parity $\frac{1}{2}^-$, because that we can multiply the interpolating currents with positive parity by the Dirac matrix $\gamma_5$ to change its parity. The hadron vacuum matrix elements in above correlation function can be defined as:
\begin{eqnarray}\label{eq:8}
\notag
\left\langle 0 \right|J_{\Lambda _c[\Xi _c]}(0)\left|\Lambda _c[\Xi _c](p') \right\rangle  &&= \lambda _{\Lambda _c[\Xi _c]}u(p',s'),\\
\notag
\left\langle 0 \right|J_{\Lambda _c[\Xi _c]}(0)\left| \Lambda _c^*[\Xi _c^*](p') \right\rangle  &&= \lambda _{\Lambda _c^*[\Xi _c^*]}\gamma _5u(p',s'),\\
\notag
\left\langle \Lambda _b[\Xi _b](p)\right|{\bar J}_{\Lambda _b[\Xi _b]}(0)\left| 0 \right\rangle  &&= \lambda _{\Lambda _b[\Xi _b]}\bar U(p,s),\\
\left\langle \Lambda _b^*[\Xi _b^*](p) \right|\bar J_{\Lambda _b[\Xi _b]}(0)\left| 0 \right\rangle  &&=  - \lambda _{\Lambda _b^*[\Xi _b^*]}\bar U(p,s)\gamma _5,
\end{eqnarray}
here $\lambda_{\Lambda^{(*)}_b[\Xi^{(*)}_b]}$ and $\lambda_{\Lambda^{(*)}_c[\Xi^{(*)}_c]}$ are the pole residues of baryons $\Lambda^{(*)}_b[\Xi^{(*)}_b]$ and $\Lambda^{(*)}_c[\Xi^{(*)}_c]$. All of these contributions must be taken into account in the calculations to obtain reliable results.

In order to make a comparison with the results of other collaborations, the following parameterization approach for form factors is used in our present work
\begin{eqnarray}\label{eq:9}
\notag
\left\langle \Lambda _c[\Xi _c](p') \right|J_\mu ^{V - A}(0)\left| \Lambda _b[\Xi _b](p) \right\rangle  &&= \bar u(p',s')\left[ F_1^{ +  + }(q^2)\frac{p^\mu }{m_{\Lambda _b[\Xi _b]}} + F_2^{ +  + }(q^2)\frac{p'^\mu }{m_{\Lambda _c[\Xi _c]}} + F_3^{ +  + }(q^2)\gamma _\mu \right]U(p,s)\\
\notag
&&- \bar u(p',s')\left[ G_1^{ +  + }(q^2)\frac{p^\mu }{m_{\Lambda _b[\Xi _b]}} + G_2^{ +  + }(q^2)\frac{p'^\mu}{m_{\Lambda _c[\Xi _c]}} + G_3^{ +  + }(q^2)\gamma _\mu \right]\gamma _5U(p,s),\\
\notag
\left\langle \Lambda _c^*[\Xi _c^*](p') \right|J_\mu ^{V - A}(0)\left| \Lambda _b[\Xi _b](p) \right\rangle  &&= \bar u(p',s')\gamma _5\left[ F_1^{ +  - }(q^2)\frac{p^\mu }{m_{\Lambda _b[\Xi _b]}} + F_2^{ +  - }(q^2)\frac{p'^\mu }{m_{\Lambda _c^*[\Xi _c^*]}} + F_3^{ +  - }(q^2)\gamma _\mu \right]U(p,s)\\
\notag
&&- \bar u(p',s')\gamma _5\left[ G_1^{ +  - }(q^2)\frac{p^\mu }{m_{\Lambda _b[\Xi _b]}} + G_2^{ +  - }(q^2)\frac{p'^\mu }{m_{\Lambda _c^*[\Xi _c^*]}} + G_3^{ +  - }(q^2)\gamma _\mu \right]\gamma _5U(p,s),\\
\notag
\left\langle \Lambda _c[\Xi _c](p') \right|J_\mu ^{V - A}(0)\left| \Lambda _b^*[\Xi _b^*](p) \right\rangle  &&= \bar u(p',s')\left[ F_1^{ -  + }(q^2)\frac{p^\mu }{m_{\Lambda _b^*[\Xi _b^*]}} + F_2^{ -  + }(q^2)\frac{p'^\mu }{m_{\Lambda _c[\Xi _c]}} + F_3^{ -  + }(q^2)\gamma _\mu \right]\gamma _5U(p,s)\\
\notag
&&- \bar u(p',s')\left[ G_1^{ -  + }(q^2)\frac{p^\mu }{m_{\Lambda _b^*[\Xi _b^*]}} + G_2^{ -  + }(q^2)\frac{p'^\mu }{m_{\Lambda _c[\Xi _c]}} + G_3^{ -  + }(q^2)\gamma _\mu \right]U(p,s),\\
\notag
\left\langle \Lambda _c^*[\Xi _c^*](p') \right|J_\mu ^{V - A}(0)\left| \Lambda _b^*[\Xi _b^*](p) \right\rangle  &&= \bar u(p',s')\gamma _5\left[ F_1^{ -  - }(q^2)\frac{p^\mu}{m_{\Lambda _b^*[\Xi _b^*]}} + F_2^{ -  - }(q^2)\frac{p'^\mu }{m_{\Lambda _c^*[\Xi _c^*]}} + F_3^{ -  - }(q^2)\gamma _\mu \right]\gamma _5U(p,s)\\
&&- \bar u(p',s')\gamma _5\left[ G_1^{ -  - }(q^2)\frac{p^\mu}{m_{\Lambda _b^*[\Xi _b^*]}} + G_2^{ -  - }(q^2)\frac{p'^\mu}{m_{\Lambda _c^*[\Xi _c^*]}} + G_3^{ -  - }(q^2)\gamma _\mu\right]U(p,s),
\end{eqnarray}
where the superscript of form factors indicates the parity of the initial and final state hadrons in transition matrix elements. For example, $+-$ and $-+$ denote the $\Lambda_b[\Xi_b]\to\Lambda^*_c[\Xi^*_c]$ and $\Lambda^*_b[\Xi^*_b]\to\Lambda_c[\Xi_c]$ transitions, respectively. Substituting the matrix elements in Eq. (\ref{eq:7}) with Eqs. (\ref{eq:8}) and (\ref{eq:9}),
the correlation functions in phenomenological side can be written as:
\begin{eqnarray}\label{eq:10}
\notag
\Pi _\mu ^{\mathrm{phy}}(p,p') &&= \frac{\lambda _{\Lambda _c[\Xi _c]}\lambda _{\Lambda _b[\Xi _b]}(\slashed p' + m_{\Lambda _c[\Xi _c]})\left\{ \begin{array}{l}
\left[ F_1^{ +  + }(q^2)\frac{p^\mu}{m_{\Lambda _b[\Xi _b]}} + F_2^{ +  + }(q^2)\frac{p'^\mu }{m_{\Lambda _c[\Xi _c]}} + F_3^{ +  + }(q^2)\gamma _\mu \right]\\
- \left[ G_1^{ +  + }(q^2)\frac{p^\mu}{m_{\Lambda _b[\Xi _b]}} + G_2^{ +  + }(q^2)\frac{p'^\mu}{m_{\Lambda _c[\Xi _c]}} + G_3^{ +  + }(q^2)\gamma _\mu \right]\gamma _5
\end{array} \right\}(\slashed p +m_{\Lambda _b[\Xi _b]})}{(m_{\Lambda _c[\Xi _c]}^2-p'^2)(m_{\Lambda _b[\Xi _b]}^2-p^2)}\\
\notag
&&+ \frac{\lambda _{\Lambda _c^*[\Xi _c^*]}\lambda _{\Lambda _b[\Xi _b]}\gamma _5(\slashed p' + m_{\Lambda _c^*[\Xi _c^*]})\left\{ \begin{array}{l}
\gamma _5\left[ F_1^{ +  - }(q^2)\frac{p^\mu }{m_{\Lambda _b[\Xi _b]}} + F_2^{ +  - }(q^2)\frac{p'^\mu}{m_{\Lambda _c^*[\Xi _c^*]}} + F_3^{ +  - }(q^2)\gamma _\mu \right]\\
- \gamma _5\left[ G_1^{ +  - }(q^2)\frac{p^\mu}{m_{\Lambda _b[\Xi _b]}} + G_2^{ +  - }(q^2)\frac{p'^\mu}{m_{\Lambda _c^*[\Xi _c^*]}} + G_3^{ +  - }(q^2)\gamma _\mu \right]\gamma _5
\end{array} \right\}(\slashed p + m_{\Lambda _b[\Xi _b]})}{(m_{\Lambda _c^*[\Xi _c^*]}^2-p'^2)(m_{\Lambda _b[\Xi _b]}^2-p^2)}\\
\notag
&&- \frac{\lambda _{\Lambda _c[\Xi _b]}\lambda _{\Lambda _b^*[\Xi _b^*]}(\slashed p' + m_{\Lambda _c[\Xi _b]})\left\{ \begin{array}{l}
\left[ F_1^{ -  + }(q^2)\frac{p^\mu}{m_{\Lambda _b^*[\Xi _b^*]}} + F_2^{ -  + }(q^2)\frac{p'^\mu}{m_{\Lambda _c[\Xi _c]}} + F_3^{ -  + }(q^2)\gamma _\mu \right]\gamma _5\\
- \left[ G_1^{ -  + }(q^2)\frac{p^\mu}{m_{\Lambda _b^*[\Xi _b^*]}} + G_2^{ -  + }(q^2)\frac{p'^\mu}{m_{\Lambda _c[\Xi _c]}} + G_3^{ -  + }(q^2)\gamma _\mu \right]
\end{array} \right\}(\slashed p + m_{\Lambda _b^*[\Xi _b^*]})\gamma _5}{(m_{\Lambda _c[\Xi _c]}^2-p'^2)(m_{\Lambda _b^*[\Xi _b^*]}^2-p^2)}\\
\notag
&&- \frac{\lambda _{\Lambda _c^*[\Xi _c^*]}\lambda _{\Lambda _b^*[\Xi _b^*]}\gamma _5(\slashed p' +m_{\Lambda _c^*[\Xi _c^*]})\left\{ \begin{array}{l}
\gamma _5\left[ F_1^{ -  - }(q^2)\frac{p^\mu }{m_{\Lambda _b^*[\Xi _b^*]}} + F_2^{ -  - }(q^2)\frac{p'^\mu}{m_{\Lambda _c^*[\Xi _c^*]}} + F_3^{ -  - }(q^2)\gamma _\mu \right]\gamma _5\\
- \gamma _5\left[ G_1^{ -  - }(q^2)\frac{p^\mu }{m_{\Lambda _b^*[\Xi _b^*]}} + G_2^{ -  - }(q^2)\frac{p'^\mu}{m_{\Lambda _c^*[\Xi _c^*]}} + G_3^{ -  - }(q^2)\gamma _\mu \right]
\end{array} \right\}(\slashed p + m_{\Lambda _b^*[\Xi _b^*]})\gamma _5}{(m_{\Lambda _c^*[\Xi _c^*]}^2-p'^2)(m_{\Lambda _b^*[\Xi _b^*]}^2-p^2)} + ...\\
\end{eqnarray}
\end{widetext}
The above correlation function can be decomposed into the following different dirac structures:
\begin{eqnarray}
\notag
\Pi^{\mathrm{phy}}_\mu(p,p')&&=\Pi^{\mathrm{phy}}_1\gamma_\mu+\Pi^{\mathrm{phy}}_2\gamma_\mu\slashed p'+\Pi^{\mathrm{phy}}_3\gamma_\mu\slashed q + \Pi^{\mathrm{phy}}_4\gamma_\mu\slashed p'\slashed q\\
\notag
&&+\Pi^{\mathrm{phy}}_5\slashed p'p'_\mu+\Pi^{\mathrm{phy}}_6\slashed p'q_\mu+\Pi^{\mathrm{phy}}_7\slashed qp'_\mu+\Pi^{\mathrm{phy}}_8\slashed qq_\mu\\
\notag
&&+\Pi^{\mathrm{phy}}_9\slashed p'\slashed qp'_\mu+\Pi^{\mathrm{phy}}_{10}\slashed p'\slashed qq_\mu+\Pi^{\mathrm{phy}}_{11}p'_\mu+\Pi^{\mathrm{phy}}_{12}q_\mu\\
\notag
&&+\Pi^{\mathrm{phy}}_{13}\gamma_\mu\gamma_5+\Pi^{\mathrm{phy}}_{14}\gamma_\mu\slashed p'\gamma_5+\Pi^{\mathrm{phy}}_{15}\gamma_\mu\slashed q\gamma_5\\
\notag
&& + \Pi^{\mathrm{phy}}_{16}\gamma_\mu\slashed p'\slashed q\gamma_5+\Pi^{\mathrm{phy}}_{17}\slashed p'\gamma_5p'_\mu+\Pi^{\mathrm{phy}}_{18}\slashed p'\gamma_5q_\mu\\
\notag
&&+\Pi^{\mathrm{phy}}_{19}\slashed q\gamma_5p'_\mu+\Pi^{\mathrm{phy}}_{20}\slashed q\gamma_5q_\mu+\Pi^{\mathrm{phy}}_{21}\slashed p'\slashed q\gamma_5p'_\mu\\
&&+\Pi^{\mathrm{phy}}_{22}\slashed p'\slashed q\gamma_5q_\mu+\Pi^{\mathrm{phy}}_{23}\gamma_5p'_\mu+\Pi^{\mathrm{phy}}_{24}\gamma_5q_\mu.
\end{eqnarray}
There are twenty four form factors in Eq. (\ref{eq:10}) which are included in these expansion coefficients $\Pi^{\mathrm{phy}}_i$ ($i=1,...,24$). $\Pi^{\mathrm{phy}}_i$ is commonly called the scalar invariant amplitude. In this work, we are only interested in the form factors of $\Lambda_b[\Xi_b]\to\Lambda_c[\Xi_c]$ transition.

\subsection{The QCD side}\label{sec3.2}
In QCD side, we bring the specific form of interpolating currents into the correlation function in Eq. (\ref{eq:5}). After doing the OPE by Wick's theorem, the correlation function in QCD side can be expressed as the following form:
\begin{eqnarray}\label{eq:12}
\notag
\Pi _\mu ^{\mathrm{QCD}}(p,p') &&= \varepsilon _{ijk}\varepsilon _{i'j'k'}\int d^4x d^4ye^{ip'x}e^{iqy}Tr\left\{ D[S]^{k'j}(x) \right.\\
\notag
&&\left. { \times \gamma _5\mathcal{C}U^{ki'T}(x)\mathcal{C}\gamma _5} \right\}C^{j'm}(x - y)\gamma _\mu (1 - \gamma _5)B^{mi}(y),\\
\end{eqnarray}
where $U[D]^{ij}(x)$, $S^{ij}(x)$ and $C[B]^{ij}(x)$ are the full propagator of $u(d)$, $s$ and $c(b)$ quarks which can be written as follows~\cite{Pascual:1984zb,Reinders:1984sr}: 
\begin{eqnarray}\label{eq:13}
\notag
U[D]^{ij}(x) &&= \frac{i\delta ^{ij}\slashed x}{2\pi ^2x^4} - \frac{\delta ^{ij}\left\langle \bar qq \right\rangle}{12} - \frac{\delta ^{ij}x^2\left\langle \bar qg_s\sigma Gq \right\rangle }{192}\\
\notag
&&- \frac{ig_sG_{\alpha \beta }^at_{ij}^a(\slashed x\sigma ^{\alpha \beta } + \sigma ^{\alpha \beta }\slashed x)}{32\pi ^2x^2} - \frac{\left\langle \bar q^j\sigma ^{\mu \nu }q^i \right\rangle \sigma _{\mu \nu }}{8} + ...,
\end{eqnarray}
\begin{eqnarray}
\notag
S^{ij}(x) &&= \frac{i\delta ^{ij}\slashed x}{2\pi ^2x^4} - \frac{\delta ^{ij}m_s}{4\pi ^2x^2} - \frac{\delta ^{ij}\left\langle \bar ss \right\rangle }{12} + \frac{i\delta ^{ij}\slashed xm_s\left\langle \bar ss \right\rangle }{48}\\
\notag
&&- \frac{\delta ^{ij}x^2\left\langle \bar sg_s\sigma Gs \right\rangle }{192} + \frac{i\delta ^{ij}x^2\slashed xm_s\left\langle \bar s{g_s}\sigma Gs \right\rangle }{1152}\\
\notag
&&- \frac{ig_sG_{\alpha \beta }^at_{ij}^a(\slashed x\sigma ^{\alpha \beta} + \sigma ^{\alpha \beta }\slashed x)}{32\pi ^2x^2} - \frac{\left\langle {\bar s^j\sigma ^{\mu \nu }s^i} \right\rangle \sigma _{\mu \nu }}{8} + ...,
\end{eqnarray}
\begin{eqnarray}
\notag
C[B]^{ij}(x) &&= \frac{i}{(2\pi )^4}\int d^4k e^{ - ikx} \left\{ \frac{\delta ^{ij}}{\slashed k - m_{c[b]}} \right.\\
\notag
&&- \frac{g_sG_{\alpha \beta }^nt_{ij}^n}{4}\frac{\sigma ^{\alpha \beta }(\slashed k + m_{c[b]}) + (\slashed k + m_{c[b]})\sigma ^{\alpha \beta }}{(k^2 - m_{c[b]}^2)^2}\\
\notag
&&- \frac{g_s^2(t^at^b)_{ij}G_{\alpha \beta }^aG_{\mu \nu }^b(f^{\alpha \beta \mu \nu } + f^{\alpha \mu \beta \nu} + f^{\alpha \mu \nu \beta })}{4(k^2 - m_{c[b]}^2)^5}\\
&&\left. { + ...} \right\},
\end{eqnarray}
Here, $\langle\bar{q}(\bar{s})g_s\sigma Gq(s)\rangle=\langle\bar{q}(\bar{s})g_s\sigma_{\mu\nu} G^{\alpha}_{\mu\nu}t^{\alpha}q(s)\rangle$, $t^\alpha=\frac{\lambda^\alpha}{2}$, $\lambda^\alpha$ ($\alpha=1,...,8$) are the Gell-Mann matrices and $f^{\alpha \beta \mu \nu }$ can be expressed as the following form:
\begin{eqnarray}
\notag
f^{\alpha \beta \mu \nu } &&= (\slashed k + m_{c[b]})\gamma ^\alpha (\slashed k + m_{c[b]})\gamma ^\beta(\slashed k + m_{c[b]})\gamma ^\mu\\
&&\times(\slashed k + m_{c[b]})\gamma ^\nu(\slashed k + m_{c[b]}).
\end{eqnarray}
The correlation function in QCD side can also be decomposed into the following form:
\begin{eqnarray}
\notag
\Pi^{\mathrm{QCD}}_{\mu}(p,p')&&=\Pi^{\mathrm{QCD}}_1\gamma_\mu+\Pi^{\mathrm{QCD}}_2\gamma_\mu\slashed p'+\Pi^{\mathrm{QCD}}_3\gamma_\mu\slashed q \\
\notag
&&+ \Pi^{\mathrm{QCD}}_4\gamma_\mu\slashed p'\slashed q+\Pi^{\mathrm{QCD}}_5\slashed p'p'_\mu+\Pi^{\mathrm{QCD}}_6\slashed p'q_\mu\\
\notag
&&+\Pi^{\mathrm{QCD}}_7\slashed qp'_\mu+\Pi^{\mathrm{QCD}}_8\slashed qq_\mu+\Pi^{\mathrm{QCD}}_9\slashed p'\slashed qp'_\mu\\
\notag
&&+\Pi^{\mathrm{QCD}}_{10}\slashed p'\slashed qq_\mu+\Pi^{\mathrm{QCD}}_{11}p'_\mu+\Pi^{\mathrm{QCD}}_{12}q_\mu\\
\notag
&&+\Pi^{\mathrm{QCD}}_{13}\gamma_\mu\gamma_5+\Pi^{\mathrm{QCD}}_{14}\gamma_\mu\slashed p'\gamma_5+\Pi^{\mathrm{QCD}}_{15}\gamma_\mu\slashed q\gamma_5 \\
\notag
&&+ \Pi^{\mathrm{QCD}}_{16}\gamma_\mu\slashed p'\slashed q\gamma_5+\Pi^{\mathrm{QCD}}_{17}\slashed p'\gamma_5p'_\mu+\Pi^{\mathrm{QCD}}_{18}\slashed p'\gamma_5q_\mu\\
\notag
&&+\Pi^{\mathrm{QCD}}_{19}\slashed q\gamma_5p'_\mu+\Pi^{\mathrm{QCD}}_{20}\slashed q\gamma_5q_\mu+\Pi^{\mathrm{QCD}}_{21}\slashed p'\slashed q\gamma_5p'_\mu\\
\notag
&&+\Pi^{\mathrm{QCD}}_{22}\slashed p'\slashed q\gamma_5q_\mu+\Pi^{\mathrm{QCD}}_{23}\gamma_5p'_\mu+\Pi^{\mathrm{QCD}}_{24}\gamma_5q_\mu,\\
\end{eqnarray}
where $\Pi^{\mathrm{QCD}}_i$ denote scalar invariant amplitudes in QCD side, and can be written as the following double dispersion integration,
\begin{eqnarray}
	\Pi _i ^{\mathrm{QCD}}(p,p') = \int\limits_{{u_{\min }}}^\infty  du \int\limits_{{s_{\min }}}^\infty  ds \frac{\rho _i ^{\mathrm{QCD}}(s,u,q^2)}{(s - p^2)(u - p'^2)}.
\end{eqnarray}
Here $\rho^{\mathrm{QCD}}_i(s,u,q^2)$ is the QCD spectral density with $s=p^2$ and $u=p'^2$. $s_{\min}$ and $u_{\min}$ are create thresholds for $\Lambda_b[\Xi_b]$ and $\Lambda_c[\Xi_c]$ which are taken as $m_b^2[(m_b+m_s)^2]$ and $m_c^2[(m_c+m_s)^2]$, respectively.

\begin{figure*}
\centering
\includegraphics[width=16cm]{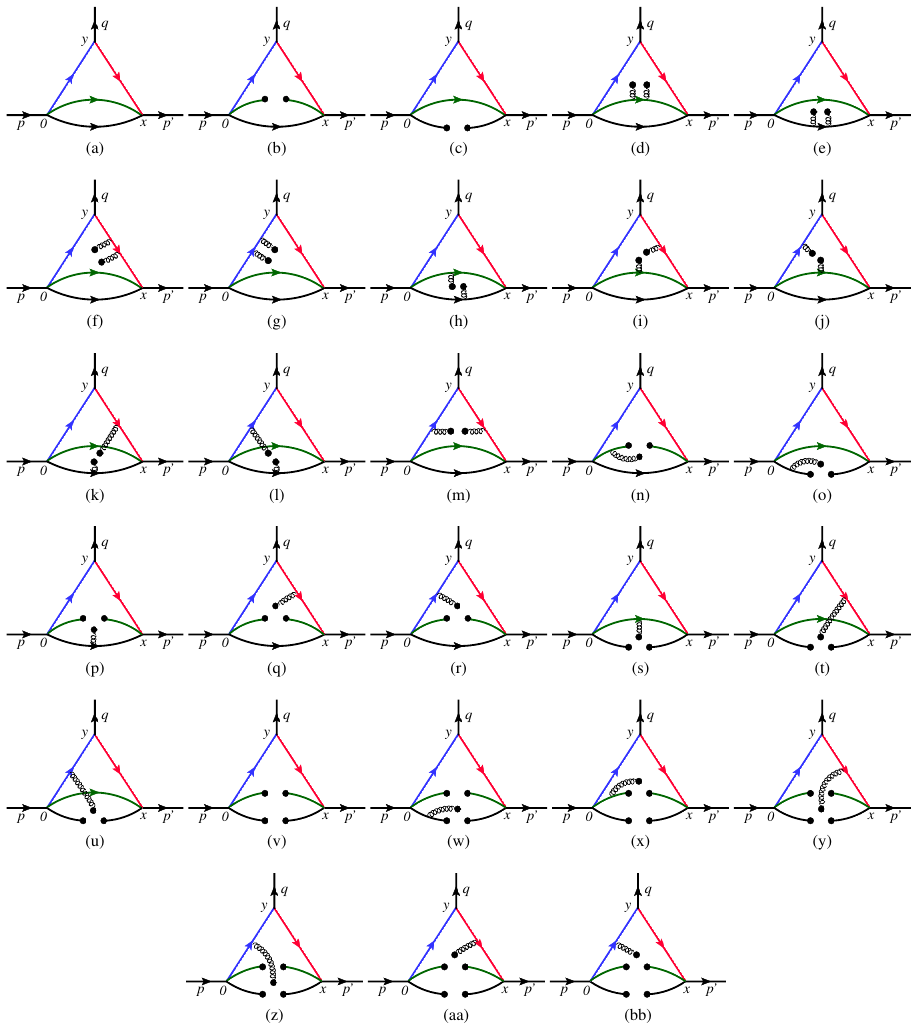}
\caption{The Feynman diagrams for the perturbative part and vacuum condensate terms, where the blue, red and black solid lines denote the $b$, $c$ and $u$ quark lines, respectively. The green solid lines denote the $d$ and $s$ quark lines for $\Lambda_b\to\Lambda_c$ and $\Xi_b\to\Xi_c$ transition, respectively. The black loop lines are the gluon lines.}
\label{FDQ}
\end{figure*}

Since the systems we studied contain light quarks, the quark condensate terms should be the dominant non-perturbative contribution. In addition, the contribution of gluon condensate can not be ignored. According our previous work~\cite{Wu:2024gcq,Lu:2023lvu}, the contributions of three gluon condensate terms are about one-tenth that of two gluon condensate terms. Thus, we only reserve the two gluon condensate terms in our calculation. The QCD spectral density can be expressed as the summation of perturbative part and different vacuum condensate terms:
\begin{eqnarray}
\notag
\rho _i ^{\mathrm{QCD}}(s,u,q^2) &&=\rho_i^{pert}(s,u,q^2)+\rho_i^{\langle \bar{q}q\rangle}(s,u,q^2)\\
\notag
&&+\rho_i^{\langle \bar{s}s\rangle}(s,u,q^2)+\rho_i^{\langle g^2_sGG\rangle}(s,u,q^2)\\
\notag
&&+\rho_i^{\langle \bar{q}g_s\sigma Gq\rangle}(s,u,q^2)+\rho_i^{\langle \bar{s}g_s\sigma Gs\rangle}(s,u,q^2)\\
\notag
&&+\rho_i^{\langle \bar{q}q\rangle^2}(s,u,q^2)+\rho_i^{\langle \bar{q}q\rangle \langle \bar{s}s\rangle}(s,u,q^2)\\
\notag
&&+\rho_i^{\langle \bar{q}q\rangle\langle \bar{q}g_s\sigma Gq\rangle}(s,u,q^2)+\rho_i^{\langle \bar{q}q\rangle\langle \bar{s}g_s\sigma Gs\rangle}(s,u,q^2)\\
&&+\rho_i^{\langle \bar{s}s\rangle\langle \bar{q}g_s\sigma Gq\rangle}(s,u,q^2),
\end{eqnarray}
where the QCD spectral density for perturbative part, the condensate terms of $\langle \bar{q}q\rangle$($\langle \bar{s}s\rangle$), $\langle g_s^2GG\rangle$ and $\langle \bar{q}g_s\sigma Gq\rangle$($\langle \bar{s}g_s\sigma Gs\rangle$) can be obtained by Cutkoskys's rules~\cite{Cutkosky:1960sp}. The calculation details can be found in Ref.~\cite{Lu:2025zaf}. For the condensate terms of $\langle \bar{q}q\rangle^2$($\langle \bar{q}q\rangle\langle \bar{s}s\rangle$) and $\langle \bar{q}q\rangle\langle \bar{q}g_s\sigma Gq\rangle$($\langle \bar{q}q\rangle\langle \bar{s}g_s\sigma Gs\rangle$ or $\langle \bar{s}s\rangle\langle \bar{q}g_s\sigma Gq\rangle$), the spectral density can be obtained directly by taking the imaginary part of the correlation function twice. For $\Lambda_b\to\Lambda_c$ transition as an example, the correlation function for these contributions can be written as:
\begin{eqnarray}
\notag
\Pi _\mu ^{\left\langle \bar qq \right\rangle^2} &&=  \frac{\left\langle \bar qq \right\rangle^2}{24}\frac{Tr\{ \gamma _5\gamma _5\} (\slashed p' + m_c)\gamma _\mu (1 - \gamma _5)(\slashed p + m_b)}{(p'^2 - m_c^2+i\epsilon_2)(p^2 - m_b^2+i\epsilon_1)},\\
\notag
\Pi _\mu ^{\langle \bar qq \rangle\langle \bar{q}g_s\sigma Gq\rangle} &&=- \frac{\langle \bar qq \rangle \langle \bar{q}g_s\sigma Gq\rangle}{48}\left[\frac{\partial}{\partial A}-m_c^2\frac{\partial^2}{(\partial A)^2}\right] \\
\notag
&&\times\frac{Tr\{ \gamma _5\gamma _5\} (\slashed p' + m_c)\gamma _\mu (1 - \gamma _5)(\slashed p + m_b)}{(p'^2 - A+i\epsilon_2)(p^2 - m_b^2+i\epsilon_1)}\big{|}_{A=m_c^2}.\\
\end{eqnarray}
The spectral density can be expressed as:
\begin{eqnarray}
\notag
&&\rho _\mu^{\langle \bar qq \rangle^2}(s,u,q^2)\\
\notag
&&= \frac{1}{\pi^2}\mathop{\rm Im}\limits_{\varepsilon _1 \to 0}\mathop{\rm Im}\limits_{\varepsilon _2 \to 0} [\Pi _\mu^{\langle \bar qq \rangle^2}(p^2+i\epsilon_1,p'^2+i\epsilon_2,q^2)],\\
\notag
&&\rho _\mu^{\langle \bar qq \rangle\langle \bar{q}g_s\sigma Gq\rangle}(s,u,q^2) \\
\notag
&&= \frac{1}{\pi^2}\mathop{\rm Im}\limits_{\varepsilon _1 \to 0}\mathop{\rm Im}\limits_{\varepsilon _2 \to 0} [\Pi _\mu^{\langle \bar qq \rangle\langle \bar{q}g_s\sigma Gq\rangle}(p^2+i\epsilon_1,p'^2+i\epsilon_2,q^2)].\\
\end{eqnarray}
After using the formula,
\begin{eqnarray}
\frac{1}{z\pm i\epsilon}=\mathcal{P}\left(\frac{1}{z}\right)\mp i\pi\delta(z)
\end{eqnarray}
The QCD spectral density for these condensate terms can be obtained as follows:
\begin{eqnarray}
\notag
\rho _\mu ^{\langle \bar qq\rangle^2}(s,u,q^2) &&= \frac{\langle \bar qq \rangle^2}{24}\delta (s - m_b^2)\delta (u - m_c^2)Tr\{ \gamma _5\gamma _5\} \\
\notag
&&\times (\slashed p' + m_c)\gamma _\mu(1 -\gamma _5)(\slashed p + m_b)\big{|}_{_{q = p - p'}^{p^2= s,p'^2 = u}},\\
\notag
\rho _\mu ^{\langle \bar qq \rangle\langle \bar{q}g_s\sigma Gq\rangle}(s,u,q^2) &&= -\frac{\langle \bar qq \rangle\langle \bar{q}g_s\sigma Gq\rangle}{48}\left[\frac{\partial}{\partial A}-m_c^2\frac{\partial^2}{(\partial A)^2}\right] \\
\notag
&&\times\delta (s - m_b^2)\delta (u - A)Tr\{ \gamma _5\gamma _5\} \\
\notag
&&\times (\slashed p' + m_c)\gamma _\mu(1 -\gamma _5)(\slashed p + m_b)\big{|}_{_{q = p - p',A=m_c^2}^{p^2= s,p'^2 = u}}.\\
\end{eqnarray}
The Feynman diagrams for perturbative part and different vacuum condensate are explicitly shown in Fig. \ref{FDQ}.
\subsection{QCD sum rules for form factors} 
Taking the variables change $p^2\to-P^2$, $p'^2\to-P'^2$ and $q^2\to-Q^2$, and performing the double Borel transforms~\cite{Reinders:1984sr} for variables $P^2$ and $P'^2$ to both phenomenological and QCD sides. The variables $P^2$ and $P'^2$ will be replaced by
Borel parameters $T_1^2$ and $T_2^2$. For simplicity, the relations $T^2=T_1^2$, $\frac{T_2^2}{T_1^2}=k$ and $k=\frac{m_{\Lambda_c[\Xi_c]}^2}{m_{\Lambda_b[\Xi_b]}^2}$~\cite{Bracco:2011pg} are adopted in our present work. Then, using the quark-hadron duality condition, we can establish a series of linear equations about twenty four scalar invariant amplitudes. Finally, all form factors in Eq. (\ref{eq:9}) can be uniquely determined by solving these twenty four linear equations. In this article, we are only concerned with the form factors for the transition from positive parity to positive parity. The QCD sum rules for corresponding form factors are as follows:
\begin{widetext}
\begin{eqnarray}\label{eq:22}
\notag
F_1^{ +  + }(Q^2) &&= \frac{m_{\Lambda _b[\Xi _b]}e^{m^2_{\Lambda _b[\Xi _b]}/T^2 + m^2_{\Lambda _c[\Xi _c]}/kT^2}}{\lambda _{\Lambda _c[\Xi _c]}{\lambda _{\Lambda _b[\Xi _b]}}(m_{\Lambda _c[\Xi _c]} + m_{\Lambda _c^*[\Xi _c^*]})(m_{\Lambda _b[\Xi _b]}+ m_{\Lambda _b^*[\Xi _b^*]})}\int\limits_{{u_{\min }}}^{u_0} du \int\limits_{{s_{\min }}}^{s_0} ds e^{ - s/T^2 - u/kT^2}\\
\notag
&&\times \left\{ {\rho _{12}^{\mathrm{QCD}}(s,u,Q^2) + m_{\Lambda _c^*[\Xi _c^*]}\rho _6^{\mathrm{QCD}}(s,u,Q^2) + (m_{\Lambda _b^*[\Xi _b^*]} - m_{\Lambda _c[\Xi _c]})[m_{\Lambda _c^*[\Xi _c^*]}\rho _{10}^{\mathrm{QCD}}(s,u,Q^2) + \rho _8^{\mathrm{QCD}}(s,u,Q^2)]} \right\},
\end{eqnarray}
\begin{eqnarray}
\notag
F_2^{ +  + }(Q^2) &&= \frac{m_{\Lambda _c[\Xi _c]}e^{m^2_{\Lambda _b[\Xi _b]}/T^2 + m^2_{\Lambda _c[\Xi _c]}/kT^2}}{\lambda _{\Lambda _c[\Xi _c]}{\lambda _{\Lambda _b[\Xi _b]}}(m_{\Lambda _c[\Xi _c]} + m_{\Lambda _c^*[\Xi _c^*]})(m_{\Lambda _b[\Xi _b]} + m_{\Lambda _b^*[\Xi _b^*]})}\int\limits_{{u_{\min }}}^{u_0} du \int\limits_{{s_{\min }}}^{s_0} ds e^{-s/T^2 - u/kT^2}\\
\notag
&&\times \left\{ \begin{array}{l}
\rho _{11}^{\mathrm{QCD}}(s,u,Q^2) - \rho _{12}^{\mathrm{QCD}}(s,u,Q^2) + 2[\rho _2^{\mathrm{QCD}}(s,u,Q^2) - \rho _3^{\mathrm{QCD}}(s,u,Q^2)]\\
+ m_{\Lambda _c^*[\Xi _c^*]}[2\rho _4^{\mathrm{QCD}}(s,u,Q^2) + \rho _5^{\mathrm{QCD}}(s,u,Q^2) - \rho _6^{\mathrm{QCD}}(s,u,Q^2)]\\
+ (m_{\Lambda _b^*[\Xi _b^*]} - m_{\Lambda _c[\Xi _c]})[m_{\Lambda _c^*[\Xi _c^*]}\rho _9^{\mathrm{QCD}}(s,u,Q^2) - m_{\Lambda _c^*[\Xi _c^*]}\rho _{10}^{\mathrm{QCD}}(s,u,Q^2) + 2\rho _4^{\mathrm{QCD}}(s,u,Q^2) + \rho _7^{\mathrm{QCD}}(s,u,Q^2)\\ - \rho _8^{\mathrm{QCD}}(s,u,Q^2)]
\end{array} \right\},
\end{eqnarray}
\begin{eqnarray}
\notag
F_3^{ +  + }(Q^2) &&= \frac{e^{m^2_{\Lambda _b[\Xi _b]}/T^2 + m^2_{\Lambda _c[\Xi _c]}/kT^2}}{\lambda _{\Lambda _c[\Xi _c]}\lambda _{\Lambda _b[\Xi _b]}(m_{\Lambda _c[\Xi _c]} + m_{\Lambda _c^*[\Xi _c^*]})(m_{\Lambda _b[\Xi _b]} + m_{\Lambda _b^*[\Xi _b^*]})}\int\limits_{{u_{\min }}}^{u_0} du \int\limits_{{s_{\min }}}^{s_0} ds e^{-s/T^2 - u/kT^2}\\
\notag
&&\times \left\{ \rho _1^{\mathrm{QCD}}(s,u,Q^2) + (m_{\Lambda _b^*[\Xi _b^*]} + m_{\Lambda _c[\Xi _c]})\rho _3^{\mathrm{QCD}}(s,u,Q^2) - m_{\Lambda _c^*[\Xi _c^*]}[\rho _2^{\mathrm{QCD}}(s,u,Q^2) + (m_{\Lambda _b^*[\Xi _b^*]} + m_{\Lambda _c[\Xi _c]})\rho _4^{\mathrm{QCD}}(s,u,Q^2)] \right\},\\
\end{eqnarray}
\begin{eqnarray}\label{eq:23}
\notag
G_1^{ +  + }(Q^2) &&= \frac{m_{\Lambda _b[\Xi _b]}e^{m^2_{\Lambda _b[\Xi _b]}/T^2 + m^2_{\Lambda _c[\Xi _c]}/kT^2}}{\lambda _{\Lambda _c[\Xi _c]}\lambda _{\Lambda _b[\Xi _b]}(m_{\Lambda _c[\Xi _c]} + m_{\Lambda _c^*[\Xi _c^*]})(m_{\Lambda _b[\Xi _b]} + m_{\Lambda _b^*[\Xi _b^*]})}\int\limits_{{u_{\min }}}^{u_0} du \int\limits_{{s_{\min }}}^{s_0} ds e^{ - s/T^2 - u/kT^2}\\
\notag
&&\times \left\{ \rho _{24}^{\mathrm{QCD}}(s,u,Q^2) + m_{\Lambda _c^*[\Xi _c^*]}\rho _{18}^{\mathrm{QCD}}(s,u,Q^2) - (m_{\Lambda _b^*[\Xi _b^*]} + m_{\Lambda _c[\Xi _c]})[m_{\Lambda _c^*[\Xi _c^*]}\rho _{22}^{\mathrm{QCD}}(s,u,Q^2) + \rho _{20}^{\mathrm{QCD}}(s,u,Q^2)] \right\},
\end{eqnarray}
\begin{eqnarray}
\notag
G_2^{ +  + }(Q^2) &&= \frac{m_{\Lambda _c[\Xi _c]}e^{m^2_{\Lambda _b[\Xi _b]}/T^2 + m^2_{\Lambda _c[\Xi _c]}/kT^2}}{\lambda _{\Lambda _c[\Xi _c]}\lambda _{\Lambda _b[\Xi _b]}(m_{\Lambda _c[\Xi _c]} + m_{\Lambda _c^*[\Xi _c^*]})(m_{\Lambda _b[\Xi _b]} + m_{\Lambda _b^*[\Xi _b^*]})}\int\limits_{{u_{\min }}}^{u_0} du \int\limits_{{s_{\min }}}^{s_0} ds e^{-s/T^2 - u/kT^2}\\
\notag
&&\times \left\{ \begin{array}{l}
\rho _{23}^{\mathrm{QCD}}(s,u,Q^2) - \rho _{24}^{\mathrm{QCD}}(s,u,Q^2) + 2[\rho _{14}^{\mathrm{QCD}}(s,u,Q^2) - \rho _{15}^{\mathrm{QCD}}(s,u,Q^2)]\\
+ m_{\Lambda _c^*[\Xi _c^*]}[2\rho _{16}^{\mathrm{QCD}}(s,u,Q^2) + \rho _{17}^{\mathrm{QCD}}(s,u,Q^2) - \rho _{18}^{\mathrm{QCD}}(s,u,Q^2)]\\
+ (m_{\Lambda _b^*[\Xi _b^*]} + m_{\Lambda _c[\Xi _c]})[m_{\Lambda _c^*[\Xi _c^*]}\rho _{22}^{\mathrm{QCD}}(s,u,Q^2) - m_{\Lambda _c^*[\Xi _c^*]}\rho _{21}^{\mathrm{QCD}}(s,u,Q^2) - 2\rho _{16}^{\mathrm{QCD}}(s,u,Q^2) - \rho _{19}^{\mathrm{QCD}}(s,u,Q^2)\\
+ \rho _{20}^{\mathrm{QCD}}(s,u,Q^2)]
\end{array} \right\},
\end{eqnarray}
\begin{eqnarray}
\notag
G_3^{ +  + }(Q^2) &&= \frac{e^{m^2_{\Lambda _b[\Xi _b]}/T^2 + m^2_{\Lambda _c[\Xi _c]}/kT^2}}{\lambda _{\Lambda _c[\Xi _c]}\lambda _{\Lambda _b[\Xi _b]}(m_{\Lambda _c[\Xi _c]} + m_{\Lambda _c^*[\Xi _c^*]})(m_{\Lambda _b[\Xi _b]} + m_{\Lambda _b^*[\Xi _b^*]})}\int\limits_{{u_{\min }}}^{u_0} du \int\limits_{{s_{\min }}}^{s_0} ds e^{ - s/T^2 - u/kT^2}\\
\notag
&&\times \left\{ {\rho _{13}^{\mathrm{QCD}}(s,u,Q^2) - (m_{\Lambda _b^*[\Xi _b^*]} - m_{\Lambda _c[\Xi _c]})\rho _{15}^{\mathrm{QCD}}(s,u,Q^2) - m_{\Lambda _c^*[\Xi _c^*]}[\rho _{14}^{\mathrm{QCD}}(s,u,Q^2) - (m_{\Lambda _b^*[\Xi _b^*]} - m_{\Lambda _c[\Xi _c]})\rho _{16}^{\mathrm{QCD}}(s,u,Q^2)]} \right\}.\\
\end{eqnarray}
\end{widetext}
Here, $s_0$ and $u_0$ are threshold parameters for initial and final state hadrons which are introduced to eliminate the contributions of higher resonances and continuum states. They often fulfill the relations $s_0=(m_{\Lambda_b[\Xi_b]}+\Delta)^2$ and $u_0=(m_{\Lambda_c[\Xi_c]}+\Delta)^2$, where $\Delta$ is the energy gap between the ground and first excited states, and take the values as $0.4-0.6$ GeV~\cite{Bracco:2011pg}. For convenience, we ignore the superscript of the form factors below and use $F_i$ and $G_i$ to represent $F_i^{++}$ and $G_i^{++}$.  
\section{Numerical results and discussions}\label{sec4}
The input parameters are all listed in Table \ref{IP}, and the masses of $u$ and $d$ quarks are ignored in this analysis. The masses of $c$, $b$ and $s$ quarks and the values of vacuum condensate are energy dependent, which can be expressed as the following forms according the renormalization group equation:
\begin{eqnarray}\label{eq:24}
\notag
m_{c[b]}(\mu ) &&= m_{c[b]}(m_{c[b]})\left[\frac{\alpha _s(\mu )}{\alpha _s(m_{c[b]})}\right]^{\frac{12}{33 - 2n_f}},\\
\notag
m_s(\mu ) &&= m_s(2 \mathrm{GeV})\left[\frac{\alpha _s(\mu )}{\alpha _s(2 \mathrm{GeV})}\right]^{\frac{12}{33 - 2n_f}},\\
\notag
\left\langle \bar qq \right\rangle (\mu ) &&= \left\langle \bar qq \right\rangle (1\mathrm{GeV})\left[\frac{\alpha _s(1\mathrm{GeV})}{\alpha _s(\mu )}\right]^{\frac{12}{33 - 2n_f}},\\
\notag
\left\langle \bar ss \right\rangle (\mu ) &&= \left\langle \bar ss \right\rangle (1\mathrm{GeV})\left[\frac{\alpha _s(1\mathrm{GeV})}{\alpha _s(\mu )}\right]^{\frac{12}{33 - 2n_f}},\\
\notag
\left\langle \bar qg_s\sigma Gq \right\rangle (\mu ) &&= \left\langle \bar qg_s\sigma Gq \right\rangle (1\mathrm{GeV})\left[\frac{\alpha _s(1\mathrm{GeV})}{\alpha _s(\mu )}\right]^{\frac{2}{33 - 2n_f}},
\end{eqnarray}
\begin{eqnarray}
\notag
\left\langle \bar sg_s\sigma Gs \right\rangle (\mu ) &&= \left\langle \bar sg_s\sigma Gs \right\rangle (1\mathrm{GeV})\left[\frac{\alpha _s(1\mathrm{GeV})}{\alpha _s(\mu )}\right]^{\frac{2}{33 - 2n_f}},\\
\notag
\alpha _s(\mu ) &&= \frac{1}{b_0t}\left[ 1 - \frac{b_1}{b_0^2}\frac{\log t}{t} \right.\\
&&\left. + \frac{b_1^2(\log ^2t - \log t - 1) + b_0b_2}{b_0^4t^2} \right],
\end{eqnarray}
where $t=\log\left(\frac{\mu^2}{\Lambda_{\mathrm{QCD}}^2}\right)$, $b_0=\frac{33-2n_f}{12\pi}$, $b_1=\frac{153-19n_f}{24\pi^2}$ and $b_2=\frac{2857-\frac{5033}{9}n_f+\frac{325}{27}n_f^2}{128\pi^3}$. $\Lambda_{\mathrm{QCD}}=213$ MeV for the quark flavors $n_f=5$ in this analysis~\cite{ParticleDataGroup:2024cfk}. The minimum subtraction masses of $c$, $b$ and $s$ quarks are taken from the Particle Data Group~\cite{ParticleDataGroup:2024cfk}, which are $m_c(m_c)=1.275\pm0.025$ GeV, $m_b(m_b)=4.18\pm0.03$ GeV and $m_s(\mu=2 \mathrm{GeV})=0.095\pm0.005$ GeV. The energy scales $\mu=2$ GeV and 2.1 GeV are adopted for $\Lambda_b$ and $\Xi_b$ transitions in this analysis, respectively.
\begin{table}[htbp]
	\begin{ruledtabular}\caption{Input parameters (IP) in this work. The values of vacuum condensate are at the energy scale $\mu=1$ GeV.}
		\label{IP}
		\begin{tabular}{c c c c }
			IP&Values(GeV) &IP&Values \\ \hline
			$m_{\Lambda_b}$&5.619~\cite{ParticleDataGroup:2024cfk}&$\lambda_{\Lambda_b}$&$0.0196\pm0.0036$ GeV$^3$~\cite{Wang:2020mxk} \\
			$m_{\Lambda_b^*}$&5.912~\cite{ParticleDataGroup:2024cfk}&$\lambda_{\Xi_b}$&$0.0223\pm0.0035$ GeV$^3$~\cite{Wang:2020mxk} \\
			$m_{\Xi_b}$&5.792~\cite{ParticleDataGroup:2024cfk}&$\lambda_{\Lambda_c}$&$0.0151\pm0.0023$ GeV$^3$~\cite{Wang:2020mxk} \\
			$m_{\Xi_b^*}$&5.960~\cite{Zhang:2008pm}&$\lambda_{\Xi_c}$&$0.0221\pm0.0035$ GeV$^3$~\cite{Wang:2020mxk} \\
			$m_{\Lambda_c}$&2.286~\cite{ParticleDataGroup:2024cfk}&$\langle \bar{q}q\rangle$&$-(0.23\pm0.01)^3$ GeV$^3$~\cite{Shifman:1978by,Reinders:1984sr}\\
			$m_{\Lambda_c^*}$&2.592~\cite{ParticleDataGroup:2024cfk}&$\langle \bar{s}s\rangle$&$(0.8\pm0.1)\langle \bar{q}q\rangle$~\cite{Shifman:1978by,Reinders:1984sr} \\
			$m_{\Xi_c}$&2.468~\cite{ParticleDataGroup:2024cfk}&$\langle \bar{q}g_s\sigma Gq\rangle$&$m_0^2\langle \bar{q}q\rangle$~\cite{Shifman:1978by,Reinders:1984sr} \\
			$m_{\Xi_c^*}$&2.790~\cite{ParticleDataGroup:2024cfk}&$\langle \bar{s}g_s\sigma Gs\rangle$&$m_0^2\langle \bar{s}s\rangle$~\cite{Shifman:1978by,Reinders:1984sr} \\
			$m_e$&$0.511\times10^{-3}$~\cite{ParticleDataGroup:2024cfk}&$m_0^2$&$0.8\pm0.1$ GeV$^2$~\cite{Shifman:1978by,Reinders:1984sr} \\
			$m_\mu$&$105.7\times10^{-3}$~\cite{ParticleDataGroup:2024cfk}&$\langle g_{s}^{2}GG\rangle$&$0.47\pm0.15$ GeV$^4$~\cite{Narison:2010cg,Narison:2011xe,Narison:2011rn} \\
			$m_\tau$&1.78~\cite{ParticleDataGroup:2024cfk}&$V_{cb}$&0.041~\cite{ParticleDataGroup:2024cfk} \\
		\end{tabular}
	\end{ruledtabular}
\end{table}

According to Eqs. (\ref{eq:22}) and (\ref{eq:23}), the results of QCD sum rules for form factors depend on the Borel parameter $T^2$, continuum threshold parameters $s_0$ and $u_0$, and the square of transition momentum $Q^2$. The values of continuum threshold parameters $s_0$ and $u_0$ have been determined by the calculations of two-point QCDSR, which are $\sqrt{s_0}=6.10\pm0.10$ ($6.30\pm0.10$) GeV for $\Lambda_b[\Xi_b]$ and $\sqrt{u_0}=2.75\pm0.10$ ($3.00\pm0.10$) GeV for $\Lambda_c[\Xi_c]$, respectively~\cite{Wang:2020mxk}. An appropriate work region of Borel parameter which is named as 'Borel platform' should be selected to obtain the final results. The form factors should have a weak Borel parameter dependency in this region. At the same time, two conditions should be satisfied which are the pole dominance and the convergence of OPE. The pole contributions for $s$ and $u$ channels can be defined as~\cite{Zhao:2020mod,Zhang:2023nxl}:
\begin{eqnarray}\label{eq:25}
\mathrm{Pole}_s=\frac{\int\limits_{u_{\min }}^{u_0} du \int\limits_{s_{\min }}^{s_0} ds}{\int\limits_{u_{\min }}^{u_0} du \int\limits_{s_{\min }}^{\infty} ds} ,
\mathrm{Pole}_u=\frac{\int\limits_{u_{\min }}^{u_0} du \int\limits_{s_{\min }}^{s_0} ds}{\int\limits_{u_{\min }}^{\infty} du \int\limits_{s_{\min }}^{s_0} ds}.
\end{eqnarray}
The condition of pole dominance requires that the pole contributions for $s$ and $u$ channel should be both larger than $40\%$. Since the form factors $F_{1,2}$ and $G_{1,2}$ are small, the definitions in Eq. (\ref{eq:25}) are ill-defined, so we only give the selection of Borel platform for $F_3$ and $G_3$ and assume that the same Borel platforms are applied to $F_{1,2}$ and $G_{1,2}$.

\begin{figure}
	\centering
	\includegraphics[width=8.5cm]{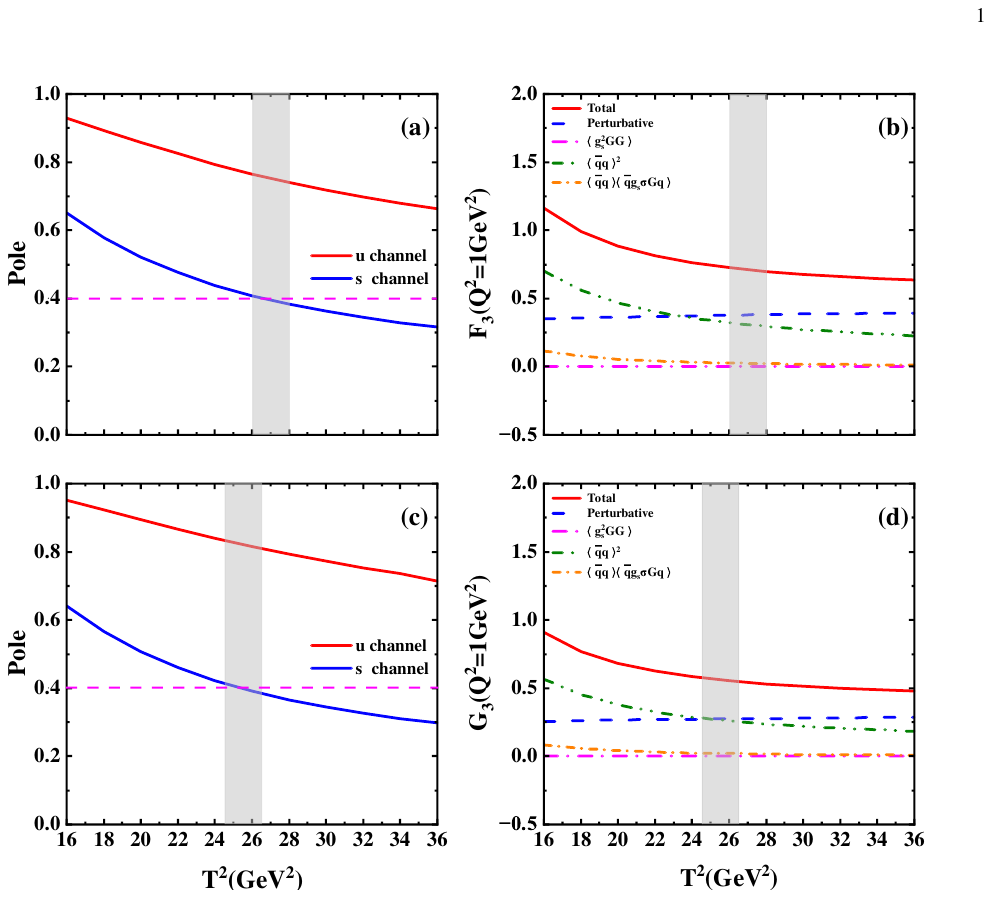}
	\caption{The pole contributions (a, c) and the contributions of perturbative term and different vacuum condensates (b, d) of form factors $F_3$ and $G_3$ for $\Lambda_b\to\Lambda_c$ transition. The grey bounds are Borel platform.}
	\label{PCandOPECL}
\end{figure}

\begin{figure}
	\centering
	\includegraphics[width=8.5cm]{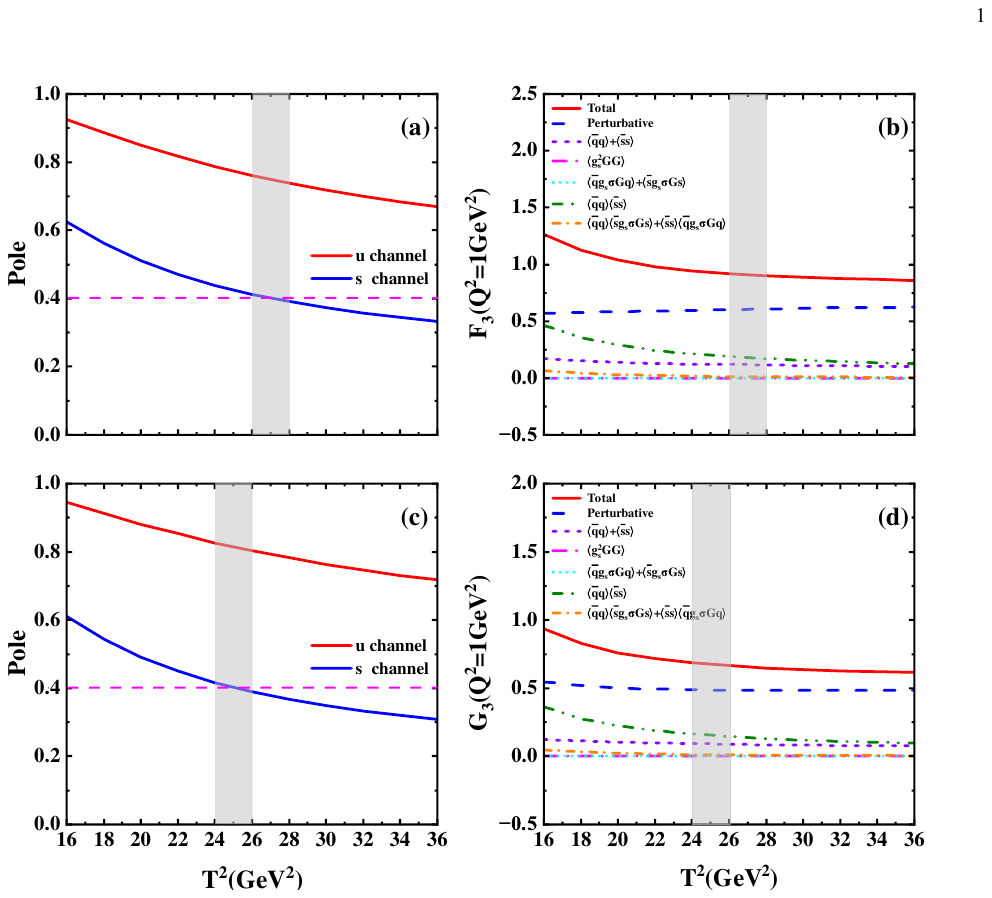}
	\caption{The pole contributions (a, c) and the contributions of perturbative term and different vacuum condensates (b, d) of form factors $F_3$ and $G_3$ for $\Xi_b\to\Xi_c$ transition. The grey bounds are Borel platform.}
	\label{PCandOPECX}
\end{figure}

Fixing $Q^2=1$ GeV$^2$, we plot the pole contribution of form factors $F_3$ and $G_3$ on Borel parameter $T^2$ (See Figs. \ref{PCandOPECL} and \ref{PCandOPECX}). Besides, the contributions of perturbative part and different vacuum condensates for $F_3$ and $G_3$ on Borel parameter $T^2$ are also shown in Figs. \ref{PCandOPECL} and \ref{PCandOPECX}. For $\Lambda_b\to\Lambda_c$ transition, the contributions from quark and quark-gluon mixed condensate are neglected because they are proportional to the mass of $u(d)$ quark. The contribution from gluon condensate is also tiny and the main contributions come from perturbative part and four quark condensate. For $\Xi_b\to\Xi_c$ transition, the contributions from gluon and quark-gluon mixed condensates are much less than $1\%$, the main contributions originate from perturbative part, quark and four quark condensates. The Borel platforms of $F_3$ and $G_3$ are determined as $26(26)-28(28)$ and $24.5(24)-26.5(26)$ GeV$^2$ for $\Lambda_b[\Xi_b]\to\Lambda_c[\Xi_c]$. The pole contributions for $s$ and $u$ channel in Borel platforms are about $40\%$. In addition, the contribution from the 8 dimension vacuum condensate term is about 4$\%$ for $\Lambda_b\to\Lambda_c$ and 2$\%$ for $\Xi_b\to\Xi_c$, this means the OPE convergence is also well satisfied. The Borel platforms, pole contributions of $s$ and $u$ channels and the contributions of perturbative part and different vacuum condensate for form factors $F_3$ and $G_3$ at $Q^2=1$ GeV$^2$ are all listed in Table \ref{PCandOPEC}.
\begin{table}[htbp]
	\begin{ruledtabular}\caption{The Borel platform (BP), pole contributions of $s$ and $u$ channel and the contributions of perturbative part and different vacuum condensate for form factors $F_3$ and $G_3$ at $Q^2=1$ GeV$^2$, where the total contributions are taken as 1.}
		\label{PCandOPEC}
		\renewcommand\arraystretch{1.3}
		\begin{tabular}{c| c c |c c}
			Mode&\multicolumn{2}{c|}{$\Lambda_b\to\Lambda_c$}&\multicolumn{2}{c}{$\Xi_b\to\Xi_c$} \\ 
			Form factor&$F_3$&$G_3$&$F_3$&$G_3$ \\ \hline
			BP (GeV$^2$)&$26-28$&$24.5-26.5$&$26-28$&$24-26$ \\
			Pole$_s$(\%)&$41-38$&$41-38$&$41-39$&$42-39$ \\
			Pole$_u$(\%)&$77-74$&$83-81$&$76-74$&$83-80$ \\
			Perturbative&$0.52-0.55$&$0.47-0.51$&$0.65-0.68$&$0.71-0.73$ \\
			$\langle\bar{q}q\rangle$&$-$&$-$&$0.13-0.12$&$0.13-0.13$ \\
			$\langle\bar{s}s\rangle$&$-$&$-$&$-$&$-$ \\
			$\langle g_s^2GG\rangle$&$\ll0.01$&$\ll0.01$&$\ll0.01$&$\ll0.01$ \\
			$\langle \bar{q}g_s\sigma Gq\rangle$&$-$&$-$&$\sim0.002$&$\sim0.002$ \\
			$\langle \bar{s}g_s\sigma Gs\rangle$&$-$&$-$&$-$&$-$ \\
			$\langle \bar{q}q\rangle^2$&$0.44-0.42$&$0.49-0.46$&$-$&$-$\\
			$\langle \bar{q}q\rangle\langle \bar{s}s\rangle$&$-$&$-$&$0.21-0.19$&$0.24-0.22$ \\
			$\langle \bar{q}q\rangle\langle \bar{q}g_s\sigma Gq\rangle$&$0.04-0.03$&$0.04-0.03$&$-$&$-$\\
			$\langle \bar{q}q\rangle\langle \bar{s}g_s\sigma Gs\rangle$&$-$&$-$&$\sim0.01$&$\sim0.01$\\
			$\langle \bar{s}s\rangle\langle \bar{q}g_s\sigma Gq\rangle$&$-$&$-$&$\sim0.01$&$\sim0.01$
		\end{tabular}
	\end{ruledtabular}
\end{table}

\begin{figure*}
	\centering
	\includegraphics[width=17cm]{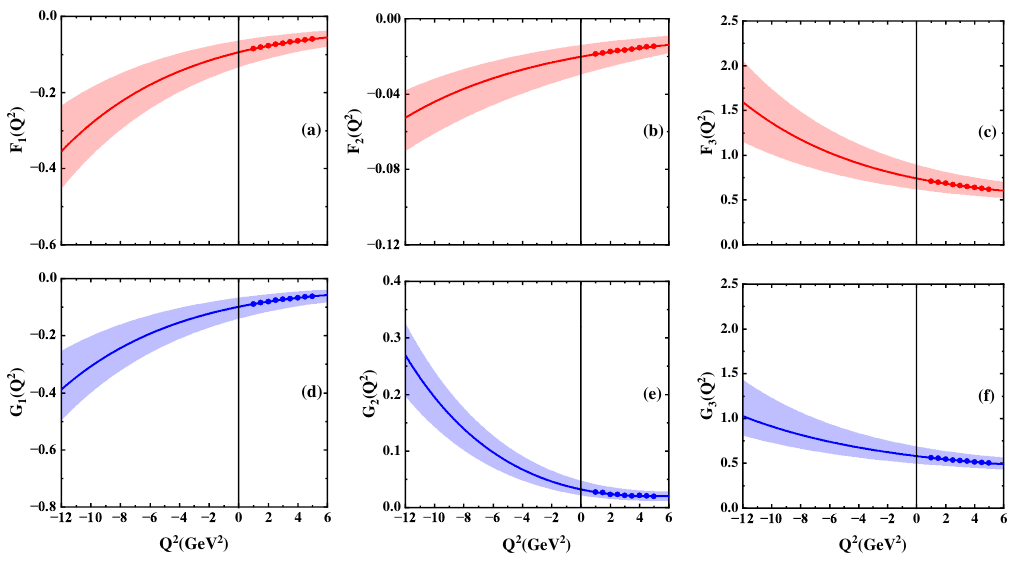}
	\caption{The fitting results of vector (a-c) and axial vector (d-f) form factors for $\Lambda_b\to\Lambda_c$ transition.}
	\label{FLQ}
\end{figure*}

\begin{figure*}
	\centering
	\includegraphics[width=17cm]{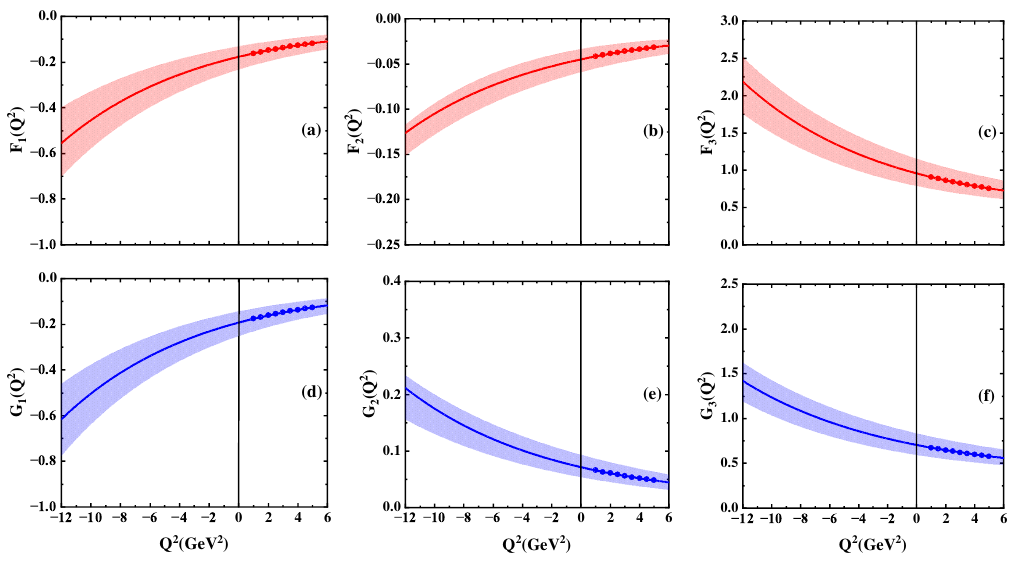}
	\caption{The fitting results of vector (a-c) and axial vector (d-f) form factors for $\Xi_b\to\Xi_c$ transition.}
	\label{FXQ}
\end{figure*}

By changing the values of $Q^2$, the form factors in space-like regions ($Q^2>0$) are obtained, where the range of $Q^2$ is taken as $1-5$ GeV$^2$. The values of these form factors in time-like regions can be obtained by fitting the results in space-like regions with an appropriate analytical function and extrapolating them into time-like regions. In present work, the $z$ series expand approach is employed to fit these form factors~\cite{Boyd:1994tt}. With this approach, the form factors can be expanded as the following series:
\begin{eqnarray}
\notag
F(Q^2) &&= \frac{1}{1 + Q^2/m_{\mathrm{pole}}^2} \\
&&\times \sum\limits_{k = 0}^{N - 1} {b_k\left[ z(Q^2,t_0)^k - {(- 1)^{k - N}}\frac{k}{N} \right.} \left. z(Q^2,t_0)^N \right],
\end{eqnarray}
where $m_{\mathrm{pole}}$ is taken as $m_{B_c}=6.275$ GeV~\cite{Zhao:2020mod}, $b_k$ is the fitting parameter, and function $z(Q^2,t_0)$ has the following form: 
\begin{eqnarray}\label{eq:28}
	z(Q^2,t_0) &&= \frac{\sqrt {t_ + + Q^2}  - \sqrt {t_ +  - t_0} }{\sqrt {t_ +  + Q^2}  + \sqrt {t_ +  - t_0}},
\end{eqnarray}
where $t_{\pm}=(m_{\Lambda_b[\Xi_b]}\pm m_{\Lambda_c[\Xi_c]})^2$, and $t_0=t_+-\sqrt{t_+(t_+-t_-)}$. The $z$ series are truncated at $N=3$ in this analysis. The values of fitting parameters $b_k$ for all form factors are listed in Table \ref{ZP} and the fitting diagrams are shown in Figs. \ref{FLQ} and \ref{FXQ}.

\begin{table}[htbp]
	\begin{ruledtabular}\caption{Fitting parameters of different form factors in $z$ series expand approach. The upper and lower bounds of the fitting parameters represent the combination of uncertainties arising from including threshold parameters, masses, pole residues of baryons, and so on, with the threshold parameters making the most significant contribution.}
		\label{ZP}
		\renewcommand\arraystretch{1.3}
		\begin{tabular}{c c c c c}
			Mode&Form factor&$b_{0}$&$b_{1}$&$b_{2}$ \\ \hline
			\multirow{6}*{$\Lambda_b \to \Lambda_c$}&$F_1$ &$-0.15^{+0.05}_{-0.05}$&$2.78^{+0.54}_{-0.97}$&$-19.59^{+7.38}_{-0.06}$ \\
			~&$F_2$ &$-0.027^{+0.008}_{-0.011}$&$0.30^{+0.06}_{-0.07}$&$-1.60^{+0.56}_{+1.45}$ \\
			~&$F_3$ &$0.86^{+0.22}_{-0.18}$&$-6.64^{+3.38}_{-3.08}$&$63.37^{+18.83}_{-29.35}$ \\
			~&$G_1$ &$-0.16^{+0.05}_{-0.06}$&$3.12^{+0.63}_{-1.13}$&$-23.05^{+9.29}_{-0.70}$ \\
			~&$G_2$ &$0.080^{+0.021}_{-0.022}$&$-2.78^{+0.71}_{-0.44}$&$32.45^{+2.46}_{-8.99}$ \\
			~&$G_3$ &$0.63^{+0.16}_{-0.12}$&$-2.41^{+1.21}_{-3.16}$&$22.82^{+30.00}_{-5.80}$ \\ \hline
			\multirow{6}*{$\Xi_b \to \Xi_c$}&$F_1$ &$-0.26^{+0.07}_{-0.07}$&$4.24^{+0.93}_{-1.28}$&$-26.91^{+10.36}_{-0.03}$ \\
			~&$F_2$ &$-0.062^{+0.012}_{-0.016}$&$0.87^{+0.09}_{-0.00}$&$-5.22^{+1.38}_{-3.86}$ \\
			~&$F_3$ &$1.17^{+0.22}_{-0.22}$&$-11.34^{+4.64}_{-0.74}$&$84.12^{+0.00}_{-17.45}$ \\
			~&$G_1$ &$-0.28^{+0.07}_{-0.08}$&$4.78^{+1.05}_{-1.20}$&$-30.28^{+8.27}_{-3.96}$ \\
			~&$G_2$ &$0.10^{+0.02}_{-0.02}$&$-1.53^{+0.41}_{-0.00}$&$8.55^{+0.00}_{-5.64}$ \\
			~&$G_3$ &$0.81^{+0.14}_{-0.13}$&$-5.70^{+0.97}_{-0.30}$&$42.35^{+0.00}_{-10.37}$ \\
		\end{tabular}
	\end{ruledtabular}
\end{table}

The values of form factors at $Q^2=0$ are listed in Table \ref{F0}. As a contrast, the results from LQCD~\cite{Detmold:2015aaa} and QCDSR~\cite{Azizi:2018axf,Zhao:2020mod,Neishabouri:2025abl} of other collaborations are also listed in this table. It can be found that the values of form factors $F_{1,2}$ and $G_{1,2}$ are close to zero in our prediction, which are consistent with the heavy quark effective theory~\cite{Manohar:2000dt}. For $\Lambda_b\to\Lambda_c$ transition, our predictions for form factors $F_3$ and $G_3$ are slightly larger than the values in Refs.~\cite{Detmold:2015aaa,Zhao:2020mod}, but are less than those in Ref.~\cite{Azizi:2018axf}. For $\Xi_b\to\Xi_c$ transition, our predictions for $F_3$ and $G_3$ are slightly larger than those in Ref.~\cite{Zhao:2020mod}. It is noted that the authors studied the semileptonic decays $\Lambda_b\to\Lambda_c$ and $\Xi_b\to\Xi_c$ by three-point QCDSR in Refs.~\cite{Azizi:2018axf,Zhao:2020mod,Neishabouri:2025abl}. However, the authors did not consider the couplings of baryon interpolating current with positive parity to negative parity baryon and did not discuss the dependence of form factors on different Dirac structures in Refs.~\cite{Azizi:2018axf,Neishabouri:2025abl}. In Ref.~\cite{Zhao:2020mod}, the OPE is only truncated at dimension of 5, the contributions of higher dimension vacuum condensates such as four quark condensate are ignored. From the Figs.~\ref{PCandOPECL} and \ref{PCandOPECX} and Table~\ref{PCandOPEC}, we can find that for the form factor $F_3$ and $G_3$, the contribution of the four quark condensate accounts for approximately 45\% of the total contribution for $\Lambda_b\to\Lambda_c$ transition and 20\% of that for $\Xi_b\to\Xi_c$ transition, respectively. This is very significant for determining the final result of the form factor.
\begin{table}[htbp]
	\begin{ruledtabular}\caption{The values of form factors (FF) at $Q^2=0$.}
		\label{F0}
		\renewcommand\arraystretch{1.3}
		\begin{tabular}{c c c c c c}
			Mode&FF&This work&\cite{Detmold:2015aaa}&\cite{Azizi:2018axf,Neishabouri:2025abl}&\cite{Zhao:2020mod} \\ \hline
			\multirow{6}*{$\Lambda_b \to \Lambda_c$}&$F_1$ &$-0.094^{+0.031}_{-0.036}$&$-0.174$&$-0.256$&$-0.101$ \\
			~&$F_2$ &$-0.021^{+0.007}_{-0.008}$&$-0.010$&$-0.421$&$-0.059$ \\
			~&$F_3$ &$0.74^{+0.15}_{-0.12}$&$0.558$&$1.220$&$0.604$ \\
			~&$G_1$ &$-0.098^{+0.028}_{-0.045}$&$-0.210$&$-0.156$&$-0.124$ \\
			~&$G_2$ &$0.032^{+0.015}_{-0.010}$&$0.082$&$0.320$&$0.080$ \\
			~&$G_3$ &$0.58^{+0.11}_{-0.09}$&$0.388$&$0.751$&$0.456$ \\ \hline
			\multirow{6}*{$\Xi_b \to \Xi_c$}&$F_1$ &$-0.18^{+0.05}_{-0.05}$&$-$&$-0.06$&$-0.125$ \\
			~&$F_2$ &$-0.045^{+0.012}_{-0.014}$&$-$&$0.28$&$-0.067$ \\
			~&$F_3$ &$0.96^{+0.20}_{-0.12}$&$-$&$0.87$&$0.701$ \\
			~&$G_1$ &$-0.19^{+0.05}_{-0.06}$&$-$&$0.30$&$-0.156$ \\
			~&$G_2$ &$0.070^{+0.020}_{-0.017}$&$-$&$-0.12$&$0.096$ \\
			~&$G_3$ &$0.71^{+0.12}_{-0.11}$&$-$&$0.53$&$0.518$ \\
		\end{tabular}
	\end{ruledtabular}
\end{table}

\begin{figure}
	\centering
	\includegraphics[width=8.5cm]{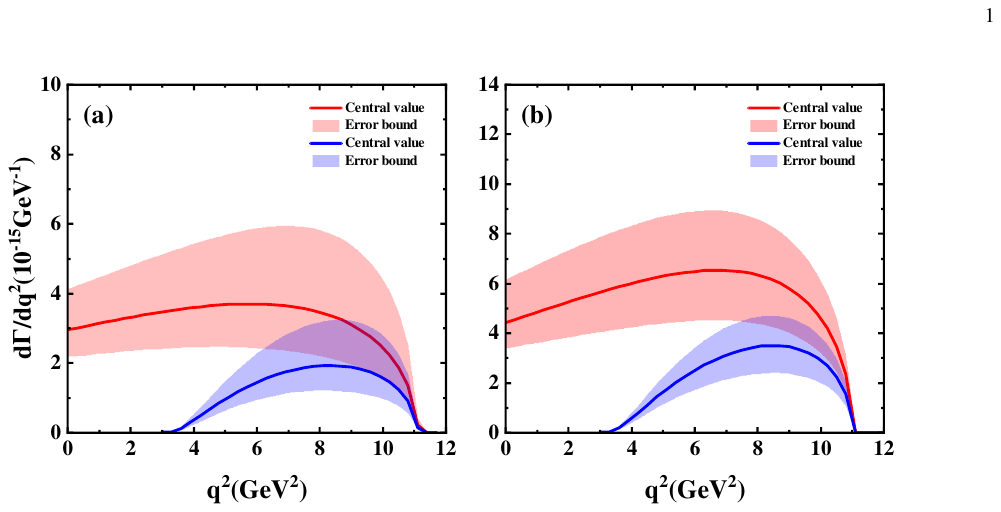}
	\caption{The differential decay width $d\Gamma/dq^{2}$ with variations of $q^{2}$ for semileptonic decay processes $\Lambda_{b}\to \Lambda_{c}l\bar{\nu}_{l}$ (a) and $\Xi_{b}\to\Xi_{c}l\bar{\nu}_{l}$ (b), where the red and blue represent $l=e$ or $\mu$ and $l=\tau$, respectively.}
	\label{DW}
\end{figure}

With the above form factors, the decay widths and branching ratios of semileptonic decays $\Lambda_b\to\Lambda_cl\bar{\nu}_l$ and $\Xi_b\to\Xi_cl\bar{\nu}_l$ can be analyzed. For the convenience of calculation, the helicity amplitudes are introduced which can be expressed as follows~\cite{Shi:2019hbf}:
\begin{eqnarray}\label{eq:39}
\notag
H_{\frac{1}{2},0}^V &&=  - i\frac{\sqrt {Q_-}}{\sqrt{q^2}}\left[ (m_{\Lambda _b[\Xi _b]} + m_{\Lambda _c[\Xi _c]})f_1(q^2) - \frac{q^2}{m_{\Lambda _b[\Xi _b]}}f_2(q^2) \right],\\
\notag
H_{\frac{1}{2},1}^V &&= i\sqrt {2Q_-} \left[  - f_1(q^2) + \frac{m_{\Lambda _b[\Xi _b]} + m_{\Lambda _c[\Xi _c]}}{m_{\Lambda _b[\Xi _b]}}f_2(q^2) \right],\\
\notag
H_{\frac{1}{2},t}^V &&=  - i\frac{\sqrt {Q_ +}}{\sqrt{q^2}}\left[ (m_{\Lambda _b[\Xi _b]} - m_{\Lambda _c[\Xi _c]})f_1(q^2) + \frac{q^2}{m_{\Lambda _b[\Xi _b]}}f_3(q^2) \right],
\end{eqnarray}
\begin{eqnarray}
\notag
H_{\frac{1}{2},0}^A &&=  - i\frac{\sqrt {Q_+}}{\sqrt{q^2}}\left[ (m_{\Lambda _b[\Xi _b]} - m_{\Lambda _c[\Xi _c]})g_1(q^2) + \frac{q^2}{m_{\Lambda _b[\Xi _b]}}g_2(q^2) \right],\\
\notag
H_{\frac{1}{2},1}^A &&= i\sqrt {2Q_+} \left[- g_1(q^2) - \frac{m_{\Lambda _b[\Xi _b]} - m_{\Lambda _c[\Xi _c]}}{m_{\Lambda _b[\Xi _b]}}g_2(q^2) \right],\\
\notag
H_{\frac{1}{2},t}^A &&=  - i\frac{\sqrt{Q_-}}{\sqrt{q^2}}\left[ (m_{\Lambda _b[\Xi _b]} + m_{\Lambda _c[\Xi _c]})g_1(q^2) - \frac{q^2}{m_{\Lambda _b[\Xi _b]}}g_3(q^2) \right], \\
\end{eqnarray}
where the superscripts $V$ and $A$ denote vector and axial vector helicity amplitudes, $Q_{\pm}=(m_{\Lambda_b[\Xi_b]}\pm m_{\Lambda_c[\Xi_c]})^2-q^2$. The negative helicity amplitudes can be obtained by
\begin{eqnarray}
\notag
H_{-\lambda _2, -\lambda _W}^V = H_{\lambda _2,\lambda _W}^V,\\
H_{-\lambda _2, -\lambda _W}^A =  - H_{\lambda _2,\lambda _W}^A,
\end{eqnarray}
where $\lambda_2$ and $\lambda_W$ represent the polarization of $\Lambda_c[\Xi_c]$ and $W$ boson, respectively. Then the total helicity amplitudes can be written as: 
\begin{eqnarray}
	H_{\lambda _2, \lambda _W}= H^V_{\lambda _2, \lambda _W}-H^A_{\lambda _2, \lambda _W}.
\end{eqnarray}
The relations of form factors $F_i$, $G_i$ and $f_i$, $g_i$ ($i=1,2,3$) are as follows:
\begin{eqnarray}
\notag
F_1(q^2)&&=f_2(q^2)+f_3(q^2),\\
\notag
F_2(q^2)&&=\frac{m_{\Lambda_c[\Xi_c]}}{m_{\Lambda_b[\Xi_b]}}[f_2(q^2)+f_3(q^2)],\\
\notag
F_3(q^2)&&=f_1(q^2)-\frac{m_{\Lambda_b[\Xi_b]}+m_{\Lambda_c[\Xi_c]}}{m_{\Lambda_b[\Xi_b]}}f_2(q^2),\\
\notag
G_1(q^2)&&=g_2(q^2)+g_3(q^2),\\
\notag
G_2(q^2)&&=\frac{m_{\Lambda_c[\Xi_c]}}{m_{\Lambda_b[\Xi_b]}}[g_2(q^2)-g_3(q^2)],\\
G_3(q^2)&&=g_1(q^2)+\frac{m_{\Lambda_b[\Xi_b]}-m_{\Lambda_c[\Xi_c]}}{m_{\Lambda_b[\Xi_b]}}g_2(q^2).
\end{eqnarray}
According to the helicity amplitudes, the differential decay width of $\Lambda_b[\Xi_b]\to\Lambda_c[\Xi_c]l\bar{\nu}_l$ can be expressed as the following forms:
\begin{eqnarray}
\notag
\frac{d\Gamma}{dq^2} &&= \frac{d\Gamma _L}{dq^2} + \frac{d\Gamma _T}{dq^2},\\
\notag
\frac{d\Gamma _L}{dq^2} &&= \frac{G_F^2V_{cb}^2q^2}{384\pi ^3m_{\Lambda _b[\Xi _b]}^2}\frac{\sqrt {Q_+Q_-}}{2m_{\Lambda _b[\Xi _b]}}{\left( 1 - \frac{m_l^2}{q^2} \right)^2}\\
\notag
&&\times \left[ \left(2 + \frac{m_l^2}{q^2}\right)\left(|H_{-\frac{1}{2},0}|^2 + |H_{\frac{1}{2},0}|^2\right) \right.\\
\notag
&&\left. { + \frac{3m_l^2}{q^2}\left(|H_{-\frac{1}{2},t}|^2 + |H_{\frac{1}{2},t}|^2\right)} \right],\\
\notag
\frac{d\Gamma _T}{dq^2} &&= \frac{G_F^2V_{cb}^2q^2}{384\pi ^3m_{\Lambda _b[\Xi _b]}^2}\frac{\sqrt{Q_+Q_-}}{2m_{\Lambda _b[\Xi _b]}}\left( 1 - \frac{m_l^2}{q^2} \right)^2\left( 2 + \frac{m_l^2}{q^2} \right)\\
&&\times \left(|H_{\frac{1}{2},1}|^2 + |H_{-\frac{1}{2}, - 1}|^2\right),
\end{eqnarray}
where $\Gamma_L$ and $\Gamma_T$ denote the longitudinally and transversely polarized decay widths. The differential decay widths with variations of $q^2$ are shown in Fig. \ref{DW}. By integrating out the square momentum $q^2$, the decay widths can be obtained as
\begin{eqnarray}
\Gamma=\int\limits_{m_l^2}^{(m_{\Lambda_b[\Xi_b]}-m_{\Lambda_c[\Xi_c]})^2}\frac{d\Gamma}{dq^2} dq^2.
\end{eqnarray}
The numerical results for decay widths and branching ratios are all collected in Table~\ref{WaB}, where the errors in the decay widths and branching ratios mainly origin from the form factors. For comparison, the experimental data and the results of other collaborations for branching ratios are also listed in this table. From the Table~\ref{WaB}, one can find that our results for the branching ratios of $\Lambda_b\to\Lambda_cl\bar{\nu}_l$ are consistent with the experimental data and the results of other collaborations within the margin of error, but the central values are slightly larger. For the branching ratios of $\Xi_b\to\Xi_cl\bar{\nu}_l$, our results are larger than those calculated using the QCDSR~\cite{Zhao:2020mod,Neishabouri:2025abl} and Light front quark model~\cite{Zhao:2018zcb}, which needs to be further verified by experiment. 

\begin{table*}[htbp]
	\renewcommand\arraystretch{1.3}
	\begin{ruledtabular}
		\caption{Decay widths (in $10^{-14}$ GeV) and Branching ratios (in \%) of $\Lambda_{b}$ and $\Xi_{b}$ semileptonic decays. Branching ratios are calculated at $\tau_{\Lambda_b^0}=(1.468\pm0.009)\times10^{-12}$ s and $\tau_{\Xi_b^0}=(1.477\pm0.032)\times10^{-12}$ s~\cite{ParticleDataGroup:2024cfk}.}
		\begin{tabular}{c >{\centering}p{6.5em}| c | c c c c c c c c}\label{WaB}
			&\multirow{2}{*}{Decay channels} & Decay widths & \multicolumn{8}{c}{Branching ratios}\\
			&&This work &This work&Experiment~\cite{ParticleDataGroup:2024cfk}&\cite{Detmold:2015aaa}&\cite{Azizi:2018axf,Neishabouri:2025abl}&\cite{Zhao:2020mod}&\cite{Gutsche:2015mxa}&\cite{Zhao:2018zcb}&\cite{Zhang:2022bvl}\\
			\hline
			&$\Lambda_{b} \to \Lambda_{c}e^-\bar{\nu}_{e}$ & $3.58^{+2.06}_{-1.18}$ & $7.98^{+4.61}_{-2.62}$ &  $6.2^{+1.4}_{-1.3}$&$5.32\pm0.35$&$6.04\pm1.70$&$6.61\pm1.08$&$6.9$&$8.83$&$6.23(58)$ \\
&$\Lambda_{b} \to \Lambda_{c}\mu^-\bar{\mu}_{e}$ & $3.56^{+2.06}_{-1.17}$& $7.95^{+4.59}_{-2.61}$ & $6.2^{+1.4}_{-1.3}$&$5.32\pm0.35$&$6.04\pm1.70$&$-$&$6.9$&$8.83$&$6.21(57)$\\
&$\Lambda_{b} \to \Lambda_{c}\tau^-\bar{\nu}_{\tau}$ & $1.04^{+0.69}_{-0.37}$ &$2.31^{+1.54}_{-0.83}$&$1.9\pm0.5$&$-$&$1.87\pm0.52$&$-$&$2.0$&$-$&$1.95(11)$ \\
&$\Xi_{b} \to \Xi_{c}e^-\bar{\nu}_{e}$ & $6.12^{+2.29}_{-1.78}$ &$13.73^{+5.14}_{-3.99}$ & $-$&$-$&$8.18^{+4.36}_{-3.34}$&$9.02\pm0.79$&$-$&$9.42$&$-$ \\
&$\Xi_{b} \to \Xi_{c}\mu^-\bar{\mu}_{e}$ & $6.10^{+2.28}_{-1.77}$ & $13.68^{+5.12}_{-3.97}$ & $-$&$-$&$8.18^{+4.36}_{-3.34}$&$-$&$-$&$9.42$&$-$ \\
&$\Xi_{b} \to \Xi_{c}\tau^-\bar{\nu}_{\tau}$ & $1.84^{+0.65}_{-0.57}$& $4.13^{+1.46}_{-1.27}$ & $-$&$-$&$2.81^{+1.50}_{-1.15}$&$-$&$-$&$-$&$-$ \\
		\end{tabular}
	\end{ruledtabular}
\end{table*}

\begin{table}[htbp]
	\renewcommand\arraystretch{1.4}
	\begin{ruledtabular}
		\caption{The lepton universality ratios (LUR) of $\Lambda_b$ and $\Xi_b$ semileptonic decays.}
		\begin{tabular}{c c c}\label{LUR}
			LUR&$R_{\Lambda_c}=\frac{\mathcal{B}[\Lambda_b\to\Lambda_c\tau\bar{\nu}_\tau]}{\mathcal{B}[\Lambda_b\to\Lambda_c\mu\bar{\nu}_\mu]}$&$R_{\Xi_c}=\frac{\mathcal{B}[\Xi_b\to\Xi_c\tau\bar{\nu}_\tau]}{\mathcal{B}[\Xi_b\to\Xi_c\mu\bar{\nu}_\mu]}$ \\ \hline
			This work&$0.29^{+0.21}_{-0.19}$&$0.30^{+0.14}_{-0.15}$ \\
			Experiment~\cite{LHCb:2017vhq,LHCb:2022piu}&$0.242\pm0.125$&$-$ \\
			LQCD~\cite{Detmold:2015aaa}&$0.3318$&$-$\\
			QCDSR~\cite{Azizi:2018axf,Neishabouri:2025abl}&$0.31\pm0.12$&$0.34\pm0.15$ \\
			CCQM~\cite{Gutsche:2015mxa}&$0.29$&$-$ \\
			RQM~\cite{Faustov:2018ahb}&$-$&$0.325$ \\
			BM~\cite{Zhang:2022bvl}&$0.31(03)$&$-$ \\
			pQCD~\cite{Rui:2025iwa,Li:2025rsm}&$0.29^{+0.12}_{-0.11}$&$0.304^{+0.023}_{-0.007}$\\
		\end{tabular}
	\end{ruledtabular}
\end{table}

The ratios of different lepton decay channels are important to test the lepton universality. Our predictions for the values of the lepton universality ratios and these of other collaboration’s are all collected in Table~\ref{LUR}. From Table~\ref{LUR}, one can find that the central value of our result for $R_{\Lambda_c}$ is close to the results predicted by other theoretical methods, including LQCD~\cite{Detmold:2015aaa}, the covariant confined quark model (CCQM)~\cite{Gutsche:2015mxa}, the bag model (BM)~\cite{Zhang:2022bvl}, and the perturbative QCD (pQCD)~\cite{Li:2025rsm}, and the predictions of all theoretical models are comparable with the experiment data. Furthermore, our prediction for $R_{\Xi_c}$ is also consistent with the results of the relativistic quark model (RQM)~\cite{Faustov:2018ahb}, pQCD~\cite{Rui:2025iwa} and previous QCDSR~\cite{Neishabouri:2025abl}.

In addition, we also analyze two asymmetry observables which are the leptonic forward-backward asymmetry $A_{FB}$ and asymmetry parameter $\alpha$. The definitions of those two asymmetry observables are as follows~\cite{Zhang:2023nxl}:
\begin{eqnarray}
	\notag
	A_{FB}(q^2)&&=\frac{d\Gamma_{\mathrm{forward}}/dq^2-d\Gamma_{\mathrm{backward}}/dq^2}{d\Gamma/dq^2},\\
	\notag
	&&=\frac{3}{4}\left({\frac{|H_{\frac{1}{2},1}|^2-|H_{-\frac{1}{2},-1}|^2}{|H_{\mathrm{tot}}|^2}}\right.\\
	\notag
	&&\left.{-\frac{2m_l^2}{q^2}\frac{H_{\frac{1}{2},0}H_{\frac{1}{2},t}^*+H_{-\frac{1}{2},0}H_{-\frac{1}{2},t}^*}{|H_{\mathrm{tot}}|^2}}\right),\\
	\alpha&&=\frac{d\Gamma_{\lambda_2=\frac{1}{2}}/dq^2-d\Gamma_{\lambda_2=-\frac{1}{2}}/dq^2}{d\Gamma_{\lambda_2=\frac{1}{2}}/dq^2+d\Gamma_{\lambda_2=-\frac{1}{2}}/dq^2},
\end{eqnarray}
where,
\begin{eqnarray}
	\notag
	|H_{\mathrm{tot}}|^2&&=\left(1+\frac{m_l^2}{2q^2}\right)\left(|H_{\frac{1}{2},1}|^2+|H_{-\frac{1}{2},-1}|^2{+|H_{\frac{1}{2},0}|^2}\right.\\
	\notag
	&&\left.{+|H_{-\frac{1}{2},0}|^2}\right)+\frac{3m_l^2}{2q^2}\left(|H_{\frac{1}{2},t}|^2+|H_{-\frac{1}{2},t}|^2\right),\\
	\notag
	\frac{d\Gamma _{\lambda_2=\pm\frac{1}{2}}}{dq^2} &&= \frac{G_F^2V_{cb}^2q^2}{384\pi ^3m_{\Lambda _b[\Xi _b]}^2}\frac{\sqrt{Q_ + Q_ - }}{2m_{\Lambda _b[\Xi _b]}}\left(1 - \frac{m_l^2}{q^2} \right)^2\\
	\notag
	&&\times \left[ \frac{4m_l^2}{3q^2}\left(|H_{\pm\frac{1}{2},1}|^2 + |H_{\pm\frac{1}{2},0}|^2+3|H_{\pm\frac{1}{2},t}|^2\right) \right.\\
	&&\left. { + \frac{8}{3}\left(|H_{\pm\frac{1}{2},1}|^2 + |H_{\pm\frac{1}{2},0}|^2\right)} \right].
\end{eqnarray}

\begin{figure}
	\centering
	\includegraphics[width=8.5cm]{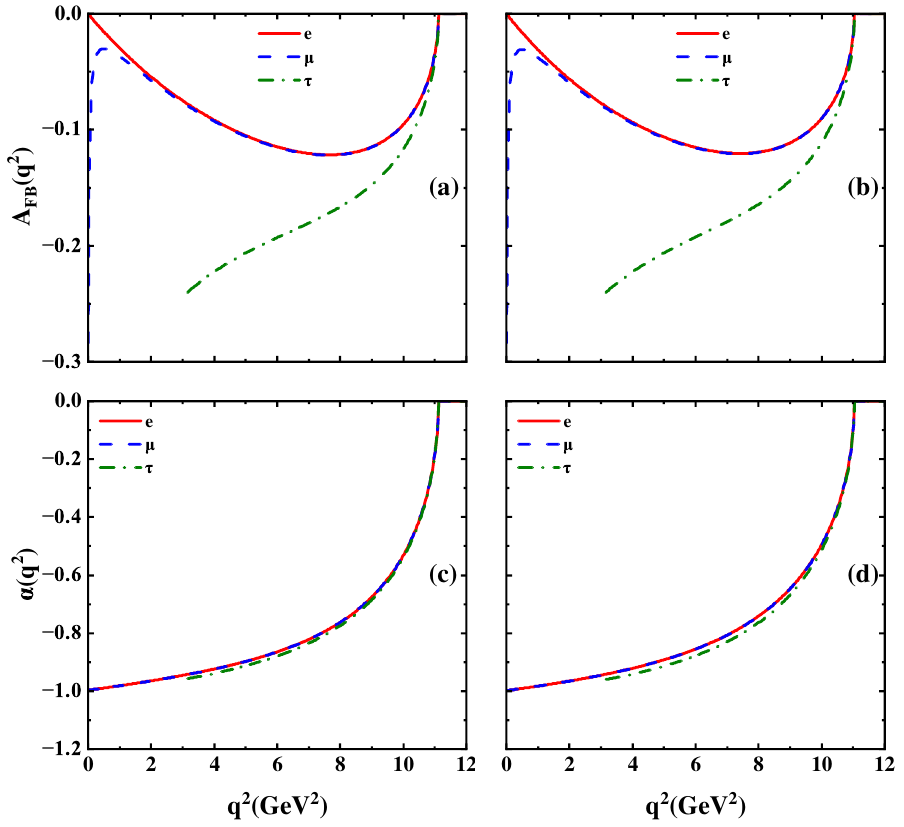}
	\caption{The leptonic forward-backward asymmetry $A_{FB}$ and asymmetry parameter $\alpha$ with variations of $q^{2}$ for decay processes $\Lambda_{b}\to \Lambda_{c}l\bar{\nu}_{l}$ (a,c) and $\Xi_{b}\to\Xi_{c}l\bar{\nu}_{l}$ (b,d), respectively.}
	\label{AFBalpha}
\end{figure}

The $q^2$ dependence of $A_{FB}$ and $\alpha$ for the decay process $\Lambda_b[\Xi_b]\to\Lambda_c[\Xi_c]l\bar{\nu}_l$ are shown in Fig.~\ref{AFBalpha}. In Figs.~\ref{AFBalpha} (a) and (b), we can see that the leptonic forward-backward asymmetry $A_{FB}$ of decay modes $\Lambda_b\to\Lambda_cl\bar{\nu}_l$ and $\Xi_b\to\Xi_cl\bar{\nu}_l$ is all going to 0 near the zero recoil point $q^2=(m_{\Lambda_b[\Xi_b]}-m_{\Lambda_c[\Xi_c]})^2$ for different leptonic decay channels. However, the behaviors of different leptonic decay channels are clearly different near the maximum recoil point $q^2=m_l^2$. The $A_{FB}$ is going to 0 and $-0.24$ for the decay processes $\Lambda_b[\Xi_b]\to\Lambda_c[\Xi_c]e\bar{\nu}_e$ and $\Lambda_b[\Xi_b]\to\Lambda_c[\Xi_c]\mu(\tau)\bar{\nu}_{\mu(\tau)}$ near $q^2=m_l^2$. As for the asymmetric parameter $\alpha$, its value increases from $-1$ to $0$ and $-0.96$ to $0$ as $q^2$ increases from $m_l^2$ to $(m_{\Lambda_b[\Xi_b]}-m_{\Lambda_c[\Xi_c]})^2$ for decay processes $\Lambda_b[\Xi_b]\to\Lambda_c[\Xi_c]e(\mu)\bar{\nu}_{e(\mu)}$ and $\Lambda_b[\Xi_b]\to\Lambda_c[\Xi_c]\tau\bar{\nu}_{\tau}$, respectively (See Figs.~\ref{AFBalpha} (c) and (d)). By finishing the integration of $q^2$ in the physical region, we calculate the average value of $A_{FB}$ and $\alpha$. For the decay process $\Lambda_b[\Xi_b]\to\Lambda_c[\Xi_c]l\bar{\nu}_l$, the central average values of $A_{FB}$ and $\alpha$ for $e$, $\mu$ and $\tau$ leptonic decay channels are given as $\langle A_{FB}\rangle_{\Lambda_c[\Xi_c]}=-0.96(-0.95)$, $-0.99(-0.98)$ and $-1.35(-1.33)$ and $\langle \alpha\rangle_{\Lambda_c[\Xi_c]}=-9.02(-8.87)$, $-9.01(-8.86)$ and $-6.04(-5.96)$, respectively. Future experiments can measure these asymmetry observables and compare them with our predictions, which will help further understand the internal structure and decay properties of single heavy baryons. Furthermore, the measurement of these asymmetry observables can also provide the possibility to explore new physics beyond the SM~\cite{Li:2021qod}.
\section{Conclusion}\label{sec5}
In this work, we firstly analyze the electroweak transition form factors of $\Lambda_b\to\Lambda_c$ and $\Xi_b\to\Xi_c$ within the framework of three-point QCDSR. In phenomenological side, all possible couplings of interpolating current to hadronic states are considered, and in QCD side, the OPE is truncated at dimension of 8. With the calculated form factors, we study the decay widths and branching ratios of semileptonic decays $\Lambda_b\to\Lambda_cl\bar{\nu}_l$ and $\Xi_b\to\Xi_cl\bar{\nu}_l$ ($l=e,\mu$ and $\tau$). Our results for the branching ratios of $\Lambda_b\to\Lambda_cl\bar{\nu}_l$ are comparable with experiment data. Moreover, the lepton universality ratio $R_{\Lambda_c[\Xi_c]}$, the leptonic forward-backward asymmetry $A_{FB}$ and the asymmetry parameter $\alpha$ are also given which can play important roles to probing potential new physics beyond the SM. Finally, we hope these results can provide useful information to research the weak decays of heavy flavor baryons in the future.

\section*{Acknowledgments}
This work is supported by National Natural Science Foundation of China under the Grant No. 12175068, as well as supported, in part, by National Key Research and Development Program under Grant No. 2024 YFA1610503 and Natural Science Foundation of HeBei Province under the Grant No. A2024502002.

\bibliography{ref.bib}
\end{document}